%% file: main.tex
\documentclass[journal]{IEEEtran}
\usepackage{amsmath,amsfonts}
\usepackage{algorithmic}
\usepackage{algorithm}
\usepackage{array}
\usepackage[caption=false,font=normalsize,labelfont=sf,textfont=sf]{subfig}
\usepackage{textcomp}
\usepackage{stfloats}
\usepackage{url}
\usepackage{verbatim}
\usepackage{graphicx}
\usepackage{cite}
\usepackage{acronym}
\usepackage{csquotes}
\usepackage{booktabs}
\usepackage{multirow}
\usepackage{nicematrix}
\usepackage{adjustbox}
\usepackage{siunitx}
\usepackage[table]{xcolor}
\usepackage{hyperref}
\usepackage[capitalize,nameinlink]{cleveref}
\crefname{equation}{}{}
\Crefname{equation}{Equation}{Equations}
\makeatletter
\newcommand*{\org@overidelabel}{}
\let\org@overridelabel\AC@verridelabel
\renewcommand*{\AC@verridelabel}[1]{%
  \@bsphack
  \protected@write\@auxout{}{\string\AC@undonewlabel{#1@cref}}%
  \org@overridelabel{#1}%
  \@esphack
}%
\makeatother

\usepackage{tikz,pgfplots}
\usepgfplotslibrary{groupplots,units}
\usetikzlibrary{patterns}
\pgfplotsset{compat=newest}
\usetikzlibrary{shapes.geometric, arrows, calc, fit, shapes, backgrounds}
\usepgfplotslibrary{colorbrewer}

\usepackage{color}
\definecolor{light}{rgb}{0.5, 0.5, 0.5}

\definecolor{answerblue}{rgb}{0.21,0.37,0.57}

\newcommand{\BE}{\begin{equation*}\begin{aligned}}
\newcommand{\EE}{\end{aligned}\end{equation*}}

\newcommand{\score}{\mathcal{A}}
\newcommand{\cX}{\mathcal{X}}

\acrodef{act}[ACT]{auxiliary classification task}
\acrodef{asd}[ASD]{anomalous sound detection}
\acrodef{cof}[COF]{connectivity-based outlier factor}
\acrodef{csls}[CSLS]{cross-domain similarity local scaling}
\acrodef{dg}[DG]{domain generalization}
\acrodef{eat}[EAT]{efficient audio transformer}
\acrodef{gmm}[GMM]{Gaussian mixture model}
\acrodef{knn}[k-NN]{k-nearest neighbors}
\acrodef{ldof}[LDOF]{local distance based outlier factor}
\acrodef{loci}[LOCI]{local correlation integral}
\acrodef{lof}[LOF]{local outlier factor}
\acrodef{loop}[LoOP]{local outlier probabilities}
\acrodef{mse}[MSE]{mean squared error}
\acrodef{ssl}[SSL]{self-supervised learning}
\acrodef{gwrp}[GWRP]{global weighted ranking pooling}
\acrodef{stft}[STFT]{short-time Fourier transform}
\acrodef{vit}[ViT]{vision transformer}
\acrodef{auc}[AUC-ROC]{area under the \acl{roc} curve}
\acrodef{pauc}[pAUC]{partial area under the \acl{roc} curve}
\acrodef{roc}[ROC]{receiver operating characteristic}
\acrodef{smote}[SMOTE]{synthetic minority over-sampling technique}

\begin{document}

\title{Local Density-Based Anomaly Score Normalization for Domain Generalization}

\author{Kevin Wilkinghoff,~\IEEEmembership{Member,~IEEE,} Haici Yang, Janek Ebbers, Fran\c{c}ois G. Germain,~\IEEEmembership{Member,~IEEE,}\\Gordon Wichern,~\IEEEmembership{Member,~IEEE,} Jonathan Le Roux,~\IEEEmembership{Fellow,~IEEE}}
        % <-this % stops a space
%\thanks{Kevin Wilkinghoff was with Mitsubishi Electric Research Laboratories (MERL), Cambridge, MA, USA. He is now with Aalborg University and Pioneer Centre for Artificial Intelligence, Aalborg, Denmark (e-mail: kevin.wilkinghoff@ieee.org).\\
%Haici Yang was an intern at MERL. She is now with Dolby Laboratories, Atlanta, GA, USA (e-mail: hy17@iu.edu).\\
%Janek Ebbers was with MERL. He is now with Amazon, Aachen, Germany (e-mail: ebbers@nt.upb.de).\\
%Fran\c{c}ois G. Germain was with MERL (e-mail: fgermain@alumni.stanford.edu).\\
%Gordon Wichern and Jonathan Le Roux are with MERL (e-mail: wichern@merl.com; leroux@merl.com).}% <-this % stops a space
%\thanks{Manuscript received April 19, 2021; revised August 16, 2021.}}



\maketitle

\begin{abstract}
State-of-the-art \ac{asd} systems in domain-shifted conditions rely on projecting audio signals into an embedding space and using distance-based outlier detection to compute anomaly scores.
One of the major difficulties to overcome is the so-called domain mismatch between the anomaly score distributions of a source domain and a target domain that differ acoustically and in terms of the amount of training data provided.
A decision threshold that is optimal for one domain may be highly sub-optimal for the other domain and vice versa.
This significantly degrades the performance when only using a single decision threshold, as is required when generalizing to multiple data domains that are possibly unseen during training while still using the same trained \ac{asd} system as in the source domain.
To reduce this mismatch between the domains, we propose a simple local-density-based anomaly score normalization scheme.
In experiments conducted on several \ac{asd} datasets, we show that the proposed normalization scheme consistently improves performance for various types of embedding-based \ac{asd} systems and yields better results than existing anomaly score normalization approaches.
\end{abstract}


\acresetall  % reset acronyms after abstract

\section{Introduction}
\IEEEPARstart{A}{nomaly} detection is the task of distinguishing between normal and anomalous data or inliers and outliers \cite{aggarwal2013outlier}.
In recent years, research in \ac{asd} has been strongly promoted by an acoustic machine condition monitoring task belonging to the annual DCASE challenge \cite{koizumi2020description,kawaguchi2021description,dohi2022description,dohi2023description,nishida2024description,nishida2025description}.
%The main challenges of this anomaly detection task are the following:
This anomaly detection task presents several difficulties.  %entails multiple challenges.
First, only normal data are assumed to be available for training. This is motivated by the fact that anomalies typically occur only rarely, and that they are very costly to obtain on purpose, as this implies damaging possibly expensive machines or waiting for them to break down; moreover, it is challenging to capture the entire diversity of all possible anomalies with a finite set of training samples.
Second, audio recordings in real-world factories tend to be very noisy. As a result, the embeddings obtained from those recordings need to be simultaneously very sensitive to changes in the target machine sound and insensitive to other acoustic events and background noise contained in the recordings, in order to reliably detect anomalies which, in comparison, may be very subtle.
Last but not least, embeddings should be easily adaptable to shifts in the acoustic environment or changes in the acoustic sources (i.e., machines), so-called domain shifts.
Ideally, users should only need to provide a very small number of samples to define how normal recordings sound like in data domains unseen during training, and a trained \ac{asd} model should still provide good results.
Such an ability is referred to as \ac{dg} \cite{wang2021generalizing,wang2023generalizing}.
Note that using target domain data for training the system was allowed in the DCASE challenge because it is impossible to tell if participants use the reference samples of the target domain for training or not.
Still, in practice it is much more favorable that systems do not need to be re-trained for every possible domain shift and thus a priori knowledge about target domains encountered during testing in the form of domain-specific training samples ideally should not be utilized for training.
An overview of methods for handling domain shifts for \ac{asd} related to DCASE can be found in \cite{wilkinghoff2025handling-eusipco}.

%%%%%%%%%%%%%%%%%%%%%%%%%%%%%%%%%%%%%%%%%%%%%%%%%%%%%%%%%%%%%%%%%%%%%%%%%%%%
\IEEEpubidadjcol  % this needs to be moved to the second column on the first page of the paper
%%%%%%%%%%%%%%%%%%%%%%%%%%%%%%%%%%%%%%%%%%%%%%%%%%%%%%%%%%%%%%%%%%%%%%%%%%%% 
\par
State-of-the-art systems for \ac{asd} are based on projecting audio signals into a relatively low-dimensional embedding space and applying general outlier detection algorithms to these embeddings afterwards \cite{wilkinghoff2024audio}.
As only normal data is available for training, training a simple binary classifier to distinguish between normal and anomalous data is impossible.
Therefore, the main difficulty when developing such \ac{asd} systems is to decide on a suitable loss function to train the embedding model that does not rely on anomalous data.
The currently best-performing embeddings are obtained by utilizing auxiliary classification tasks based on meta information such as machine types, machine IDs, or settings \cite{giri2020self,primus2020anomalous,wilkinghoff2021sub,wilkinghoff2021combining,venkatesh2022improved,germain2023hyperbolic}, or on \ac{ssl} \cite{inoue2020detection,lopez2020speaker,nejjar2022dg-mix,chen2023effective,wilkinghoff2024self}.
This auxiliary classification approach is also called outlier exposure \cite{hendrycks2019deep} as samples belonging to other classes are used as proxy outliers \cite{primus2020anomalous}.
Compared to one-class models such as autoencoders \cite{koizumi2019unsupervised,suefusa2020anomalous,giri2020group,kapka2020id,wichern2021anomalous} that treat all signal components as equally important, models trained with an auxiliary classification task learn to closely monitor target signals and ignore background noise as well as other signals as long as they do not contain useful information to solve the classification task \cite{wilkinghoff2024why}.
\par
This work, which extends our prior work on normalizing embedding-based anomaly scores for \ac{dg} \cite{wilkinghoff2024keeping}, makes several new contributions.
First and foremost, the evaluation of our proposed normalization approach is extended from a single to multiple state-of-the-art embeddings based on pre-trained models.
Furthermore, additional existing normalization approaches are reviewed, discussed, and experimentally compared to our normalization approach.
Last but not least, we provide the source code\footnote{\url{https://github.com/merlresearch/anomaly-score-normalization}} for the conducted experiments.
\par
The remaining parts of this paper are organized as follows.
In \cref{sec:related_work}, related literature for computing embedding-based anomaly scores and normalizing them is discussed.
In \cref{sec:methodology}, the proposed normalization approach is motivated, presented, and illustrated.
The effectiveness of the proposed normalization scheme is experimentally evaluated in \cref{sec:results} using the setup from \cref{sec:setup} with several state-of-the-art embeddings on multiple datasets.
In addition, a few ablation studies are carried out.
The paper is concluded with a summary and possible extensions for future work in \cref{sec:conclusions}.

\section{Related work}
\label{sec:related_work}
Anomaly scores for embedding-based \ac{asd} can be computed in several ways.
The general idea is to project the data into an embedding space obtained by training a neural network and to apply general outlier detection approaches to these embeddings \cite{schubert2014local}.
The underlying assumption is that anomalous test samples should substantially differ from normal test and training samples in the embedding space.
Among these approaches for computing anomaly scores are distance-based approaches such as \ac{knn} \cite{angiulli2002fast,deng2022ensemble,nejjar2022dg-mix}, and (global) density-based approaches based on \acp{gmm} \cite{purohit2020deep,wilkinghoff2021sub,kuroyanagi2022improvement} or the Mahalanobis distance \cite{deng2022ensemble}.
In \cite{wilkinghoff2023design}, distance-based approaches were shown to lead to better performance than density-based approaches in domain-shifted conditions because it is difficult to accurately estimate the density of the distribution in sparsely-represented target domains.
The choice of the distance function depends on the loss function of the embedding model.
For commonly used angular margin losses such as ArcFace \cite{deng2019arcface}, AdaCos \cite{zhang2019adacos}, sub-cluster AdaCos \cite{wilkinghoff2021sub}, or AdaProj \cite{wilkinghoff2024adaproj}, the cosine distance is the most natural choice to measure the distance between samples.
For other loss functions that do not act on the unit sphere, other distances such as the Euclidean distance can be used instead.
Still, distance-based approaches are sensitive to data domains with different densities, which degrades the performance (see \cref{subsec:motivation}).
Note that this negative effect can even occur when only data belonging to the source domain are encountered if the data distribution consists of clusters with varying densities.
Our proposed normalization approach is less sensitive to distributions with clusters of different densities as it also takes \emph{local} density estimates into account.
\par
There are several existing outlier detection approaches based on local density estimates. 
Their most prominent representative is the \ac{lof} \cite{breunig2000lof}, which compares the local density of a test sample to that of its nearest neighbors from a reference set.
There are multiple variants and extensions of \ac{lof}.
In \cite{papadimitriou2003loci}, the neighborhoods are defined by balls with a certain radius instead of taking the nearest neighbors of each sample.
\Ac{cof} \cite{tang2002enhancing} focuses on effectively handling low-density regions by introducing paths between nearest neighbors and utilizing decreasingly weighted lengths of the piece-wise path edges.
Other variants of \ac{lof} aim to normalize scores to increase interpretability and to simplify the setting of a decision threshold.
\Ac{ldof} \cite{zhang2009new} takes the mean distance to the $K$ nearest neighbors and normalizes it with the average pairwise distance between the $K$ nearest neighbors.
\Ac{loop} \cite{kriegel2009loop} tries to normalize outlier scores such that different outlier score distributions are similarly scaled and can be directly interpreted as probabilities.
This is achieved by introducing a so-called probabilistic distance of a query sample to a reference set. This probabilistic distance allows errors by defining spheres around the reference sample that contain all elements of a reference set only with a certain probability. Using the expected value of this distance, a probabilistic \ac{lof} is derived, whose standard deviation is used to normalize the resulting outlier scores.
The anomaly scores of \ac{lof}-based outlier detection methods take the local density into account but are based on the distance in an embedding space of a test sample to the reference samples that are closest to it according to that same distance.
Since different densities have a strong effect on the magnitude of the distance, the performance of these methods is still substantially degraded in domain-shifted conditions.
In contrast, our proposed normalization scheme first alters the anomaly scores based on the local density and only then selects the closest neighbors.
This reduces unwanted effects caused by clusters of reference samples with different densities and thus leads to anomaly scores that are more robust to domain shifts.
\par
Normalizing scores is a well-investigated topic for various related applications.
Apart from specific score normalization approaches for \ac{lof}, there are also works on normalizing outlier scores of arbitrary models such that the scores are contained in $[0,1]$ and can be interpreted as probabilities \cite{kriegel2011interpreting}.
Another example is to apply an additive similarity normalization \cite{pizzi2022self,bralios2024generation}, based on \cite{jegou2011exploiting}, by normalizing individual distance-based outlier scores with the mean outlier score of the $K$ nearest neighbors.
Speaker verification is an additional application where score normalization approaches are frequently applied, under the name of score calibration \cite{cumani2011comparison,karam2011towards}.
In contrast to our presented approach, these normalization approaches are not specifically designed for embedding-based \ac{asd} in domain-shifted conditions.
However, there are also several works that reduce the domain mismatch by normalizing the scores in domain-shifted conditions.
An example is a domain-wise standardization of the anomaly score distributions with the goal of aligning them \cite{saengthong2024deep}.
Another example is \ac{csls}, which uses two different additive terms derived from samples of the source and target domains \cite{lample2018word}.
As we will show in \cref{subsubsec:comparison}, our proposed anomaly score normalization approach leads to slightly better performance than the existing approaches while not requiring domain labels or specific training for each target domain, and allowing for independent evaluation of each test sample.

\section{Methodology}
\label{sec:methodology}
\subsection{Motivation}
\label{subsec:motivation}
\begin{figure}
    \centering
    \begin{adjustbox}{max width=\columnwidth}
          \includegraphics{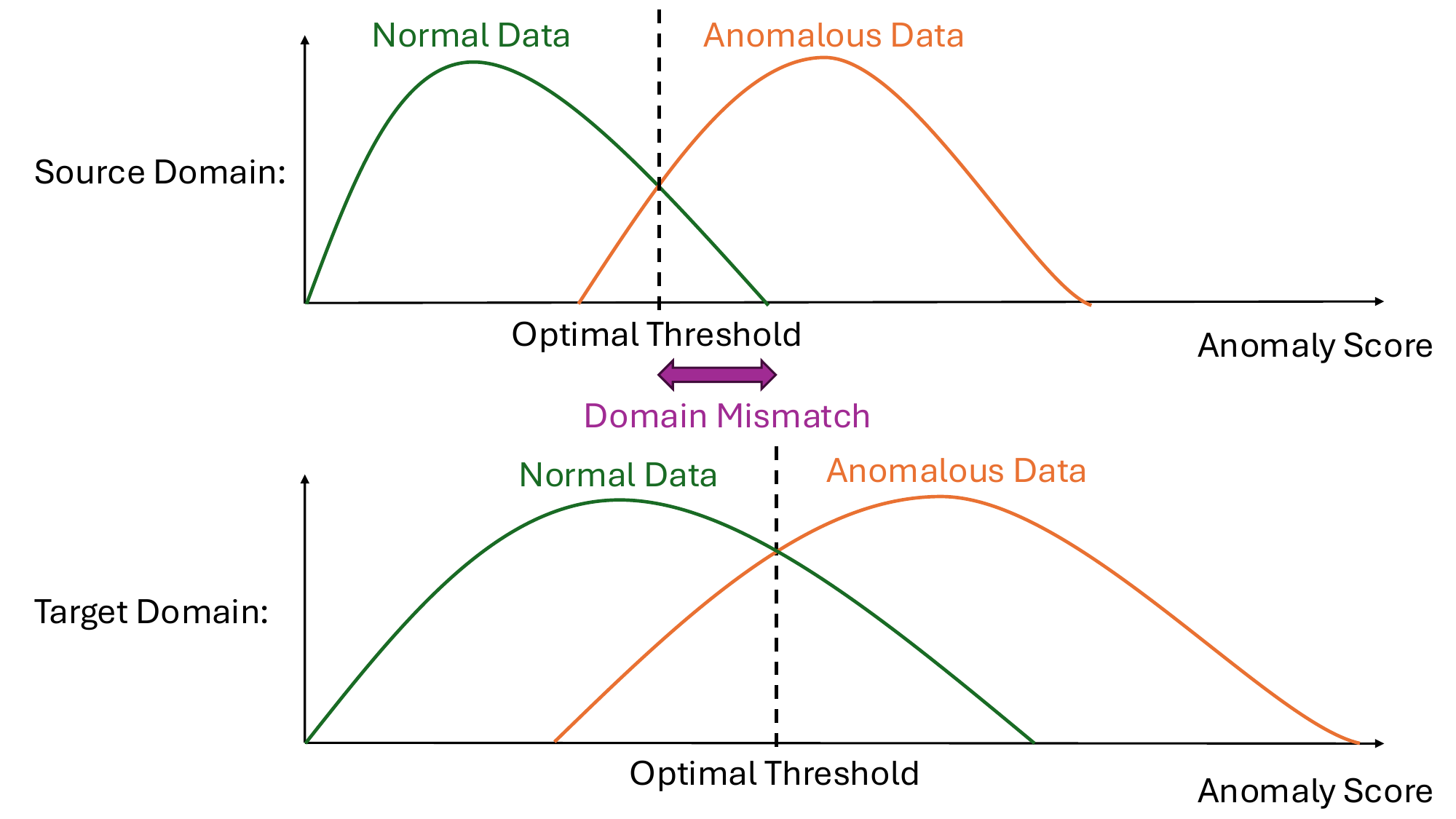}
    \end{adjustbox}
    \caption{Illustration of the domain mismatch between the anomaly scores of a source domain and a target domain. Normal and anomalous samples are usually less well separated in the target domain than in the source domain, which decreases domain-independent performance over the performance obtained for the source domain alone. Furthermore, the optimal decision thresholds for separating the scores belonging to the normal and anomalous data of different data domains differ substantially, which significantly decreases performance when using only a single threshold for both domains. Figure taken from \cite{wilkinghoff2025handling-eusipco}.}
    \label{fig:domain_mismatch}
\end{figure}
Domain shifts significantly decrease the performance of a trained \ac{asd} system because the distributions of the anomaly scores before and after the domain shift, i.e., in the source and target domains, are not necessarily well-aligned.
This makes it difficult to separate normal and anomalous samples in both domains with a single decision threshold.
This misalignment, referred to as domain mismatch, is illustrated in \cref{fig:domain_mismatch}.
Its main cause is that embedding models usually try to distribute the data into very compact clusters but are only trained with source domain data.
Note that even if target domain samples are used for training, the effect is typically negligible due to the strong imbalance of the number of available training samples between the source and target domains.
If the same embedding model is used to project data belonging to the target domain into the embedding space, the resulting clusters may have densities completely different from those of the source domain data.
Therefore, the anomaly scores may also be scaled differently.
Moreover, usually only a few reference samples are provided for the target domain, making it difficult to accurately estimate the distribution for density-based outlier detection approaches and leading to higher distances in general for distance-based approaches.
Here, clusters with different densities correspond to sub-classes of the normal data.
The main idea in this work is to reduce the domain mismatch caused by differently scaled anomaly scores.
This is achieved by normalizing the scores so that the distances between samples from both domains are more uniformly distributed.

\subsection{Distance-based anomaly score approaches}
In this section, the proposed anomaly score normalization approach will be presented.
To this end, we first introduce the notation and a simple distance-based baseline approach.
Then, two variants of the score normalization scheme will be defined.
\subsubsection{Baseline approach}
\label{subsec:baseline}
Let us now properly define the baseline approach.
Let $\cX_\text{test}$ denote the set of test samples and $\cX_\text{ref}$ denote a reference set of normal training samples that can belong to any or multiple data domains.
Following best practices, all available training samples are used as reference samples.
Then, the baseline approach for calculating an anomaly score based only on the nearest neighbor in the reference set is defined as
\BE\score^\text{NN}_\text{cos}(x, \cX_\text{ref})&:=\min_{y\in \cX_\text{ref}}\score_\text{cos}(x, y)\\&:=\min_{y\in \cX_\text{ref}}0.5\cdot\bigg(1-\frac{\langle x,y\rangle}{\lVert x\rVert_2\lVert y\rVert_2}\bigg)\in[0,1].\EE
Note that, on the unit sphere, this equation simplifies to
\BE{\score^\text{NN}_\text{cos}(x,\cX_\text{ref})=\min_{y\in \cX_\text{ref}}0.5\cdot(1-\langle x,y\rangle)}.\EE
Samples that are outliers and thus not very representative of the underlying distribution of normal samples cause unwanted effects for this distance-based anomaly score calculation approach. These can be reduced by using \ac{knn} instead of only measuring the distance to the closest neighbor \cite{deng2022ensemble,nejjar2022dg-mix}, or applying k-means to the reference set before computing the distance \cite{wilkinghoff2023design}, which also reduces computational cost at inference time.

\subsubsection{Proposed normalization approach}
\label{subsec:proposed_approach}
As mentioned in \cref{sec:related_work}, a distance-based baseline approach outperforms density-based approaches in domain-shifted conditions.
However, to achieve good performance with a distance-based approach, the anomaly scores, i.e., the distances of test samples to their closest reference samples, need to follow a similar distribution for source and target domains.
%For a distance-based approach, similarly distributed embeddings for both domains in terms of distance of an arbitrary test sample to their closest reference samples are assumed.
In most cases, this assumption is not true, causing a domain mismatch as explained in \cref{subsec:motivation}.
To reduce this mismatch, we propose two different approaches for normalizing the anomaly scores.
Both approaches are based on the local density of the reference samples and differ only in how the local density is defined.
For the first approach, the local density is based on the $K$ nearest neighbors within the reference set.
The local density of the second approach is defined by
\ac{gwrp} \cite{kolesnikov2016seed}, which was also used in \cite{guan2023time}.
\Ac{gwrp} calculates the density with all reference samples but uses an exponentially decreasing weight based on the distance-based ranking of the reference samples to put higher emphasis on closer samples and thus somehow enforce a locality constraint.
The idea for both approaches is to increase the scores measured against reference samples within regions with high density of embeddings while reducing the scores measured against samples within regions with low density.
\par
Using the notation introduced for the baseline approach in \cref{subsec:baseline}, let $y\in X_\text{ref}$ denote an element of $X_\text{ref}$, and $y_k\neq y$ denote the $k$-th closest sample in $\cX_\text{ref}$ to $y$.
Then, the normalized anomaly scores are defined as
\BE
\score^\text{K-NN}_\text{scaled}(x,\cX_\text{ref}\,|\,K)&:=\min_{y\in \cX_\text{ref}}\frac{\score_\text{cos}(x, y)}{\displaystyle \sum_{k=1}^{K}\score_\text{cos}(y,y_k)}\in\mathbb{R}_+,\\ 
\score^\text{GWRP}_\text{scaled}(x,\cX_\text{ref}\,|\,r)&:=\min_{y\in \cX_\text{ref}}\frac{\score_\text{cos}(x, y)}{\displaystyle \sum_{k=1}^{\vert \cX_\text{ref}\rvert-1}\score_\text{cos}(y,y_k)\cdot r^{k-1}}\in\mathbb{R}_+,
\EE
where the hyperparameters $K\in\mathbb{N}^+$ and $r\in[0,1]$ denote the number of %next 
closest samples to consider and the weight factor, respectively.
In our experiments, we also consider a closely-related variant where the ratio is replaced with a difference, i.e., the normalization term is subtracted instead.
It shall be emphasized that the normalization constants for all reference samples do not depend on the test sample and thus can be pre-computed, which ensures that there is no computational overhead at inference.

\subsection{Discussion of the normalization effect}
\begin{figure}
    \centering
    \begin{adjustbox}{max width=\columnwidth}
          \includegraphics{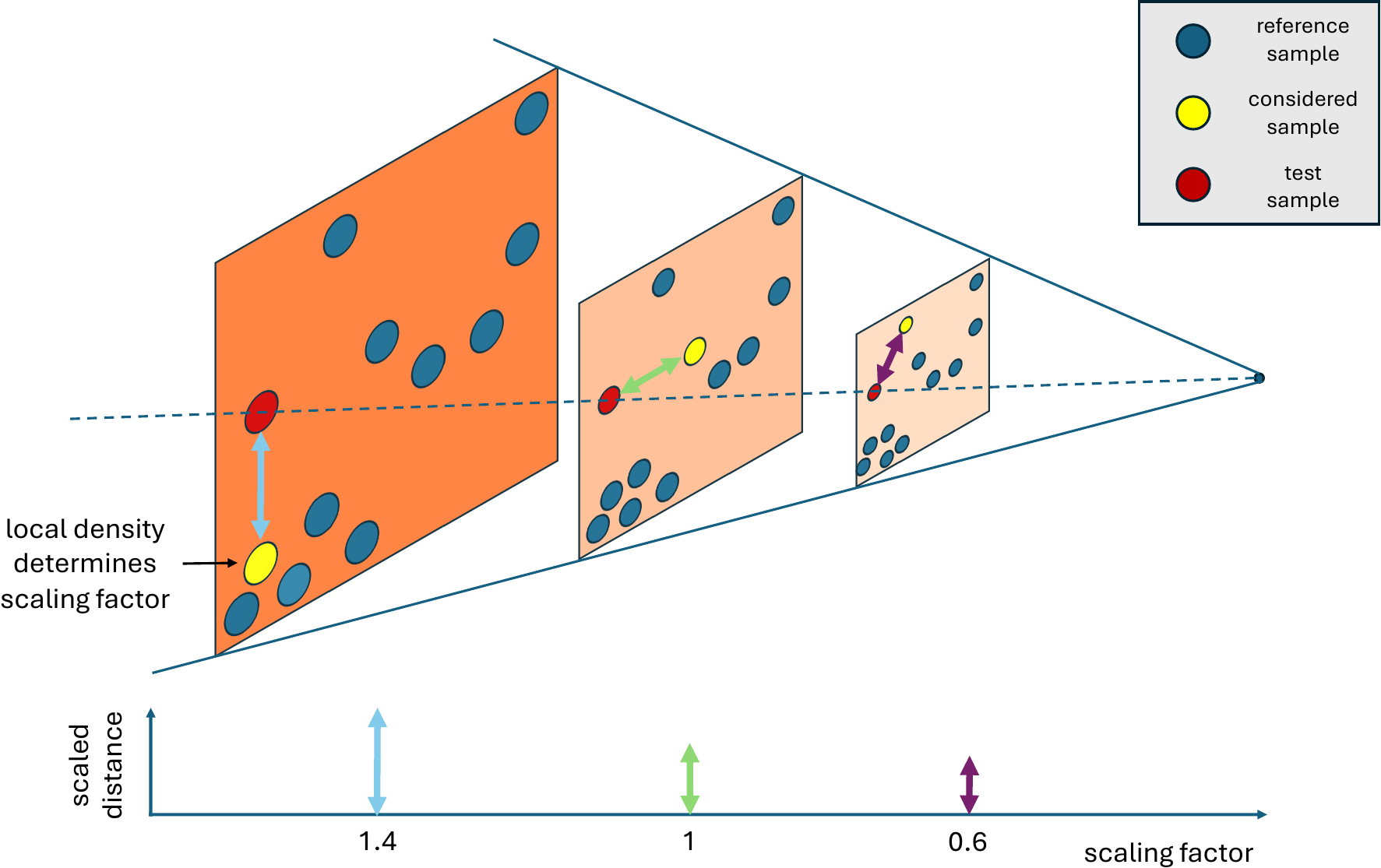}
    \end{adjustbox}
    \caption{Illustration of the impact of the ratio-based normalization approach on the selection of the reference points (in blue) that are to be considered the nearest for a given test point (in red).
    For three different considered points (in yellow) with similar distance to the test sample in the original embedding space, the scale at which they are compared to other points is shown.
    For one of them, the scaling factor is $1.4$ because the point is in a dense neighborhood; for another one, the scaling factor is $1$; and for the third sample in the sparse area, the scaling factor is $0.6$.
    When assessing the distance between the test point and any of the considered samples, the distances should be computed in the corresponding rescaled planes.
    In the end, the reference point with the smallest scaled distance is selected.
    Here, the point in the sparse area with $0.6$ scaling factor gets selected, despite the fact that all points initially had a similar distance to the test sample.
    For illustration purposes, planes are depicted here, although the normalization approach involves cosine distances on a sphere.
    For the difference-based normalization approach, reference samples are shifted based on their local densities instead of re-scaling the embedding space, which has a similar effect on the normalized distances but is more difficult to illustrate.}
    \label{fig:normalization_effect}
\end{figure}
%We will now discuss the effect of the proposed normalization scheme.
As stated before, the main goal of applying the normalization is to reduce the domain mismatch by having a more uniformly distributed embedding space across multiple domains and possibly existing sub-classes.
Effectively, this is achieved by reducing the distance of reference samples with low local density to a given test sample compared to the distance to reference samples with high density.
Therefore, if two reference samples are similarly close in the original embedding space, we favor the sample which is more isolated, i.e., the one that likely belongs to the target domain.
Still, the distance to the reference sample itself is taken into account, ensuring that after normalization the closest neighbor is a reasonable sample and not just a random outlier.
If we interpret the normalization as distorting the geometric distances from the test sample, this results in pushing reference samples with dense local neighborhoods away, while pulling samples with less dense local neighborhoods closer.
This is illustrated in \cref{fig:normalization_effect}.
As a result, the proposed normalization allows one to use a single decision threshold for samples in dense and sparse regions.
These parameters control the degree of locality, ranging from $K=1$ or $r=0$, where the local density is based only on the nearest neighbor, to a global density $K=\lvert X_\text{ref}\rvert-1$ or $r=1$, for which all other reference samples are considered.
Note that for these edge cases the parameterizations with $K$ and $r$ are the same.

% \section{Experimental evaluations}
% \label{sec:experiments}

\section{Experimental setup}
\label{sec:setup}
\subsection{Datasets}
\input{figs/datasets.tex}

For the experimental evaluations of this work, the following five datasets were used: 1) the DCASE2020 \ac{asd} dataset \cite{koizumi2020description} based on MIMII \cite{purohit2019mimii} and ToyADMOS \cite{koizumi2019toyadmos}, 2) the DCASE2022 \ac{asd} dataset \cite{dohi2022description} based on MIMII-DG \cite{dohi2022mimiidg} and ToyADMOS2 \cite{harada2021toyadmos2}, 3) the DCASE2023 \ac{asd} dataset \cite{dohi2023description} based on MIMII-DG \cite{dohi2022mimiidg} and ToyADMOS2+ \cite{harada2023toyadmos2+}, 4) the DCASE2024 \ac{asd} dataset \cite{nishida2024description} based on \cite{dohi2022mimiidg}, ToyADMOS2\# \cite{niizumi2024toyadmos2sharp}, and additional samples recorded with the same setup as presented in IMAD-DS \cite{albertini2024imadds}, and 5) the DCASE2025 \ac{asd} dataset \cite{nishida2024description} based on \cite{dohi2022mimiidg}, ToyADMOS2025 \cite{harada2025toyadmos2025}, and additional samples recorded with the same setup as presented in IMAD-DS \cite{albertini2024imadds}.
All these datasets focus on semi-supervised \ac{asd} for acoustic machine condition monitoring and consist of a development set and an evaluation set containing recordings of machines with real factory background noise.
Each development and evaluation set is divided into a training split containing only normal data and a test split containing a mix of normal and anomalous data.
A summary of the datasets can be found in \cref{tab:datasets} and more details can be found in the corresponding references.
The main differences between the datasets will now be discussed.
\par
\textbf{DCASE2020:} The DCASE 2020 \ac{asd} dataset consists of recordings belonging to six different machine types, namely \enquote{fan}, \enquote{pump}, \enquote{slider}, \enquote{valve} from MIMII \cite{purohit2019mimii}, and \enquote{ToyCar} and \enquote{ToyConveyor} from ToyAdmos \cite{koizumi2019toyadmos}.
For each machine type, there are six to seven specific machines in total, of which three to four belong to the development set and the remaining ones belong to the evaluation set.
There are approximately $1000$ normal training samples and around $400$ test samples for each individual machine.
Each recording has a length of \SI{10}{\second} and a sampling rate of \SI{16}{\kilo\hertz}.
In contrast to the other two datasets, the DCASE2020 \ac{asd} dataset does not contain any data in domain shifted conditions, i.e., it essentially only consists of a source domain.
\par
\textbf{DCASE2022:} The DCASE2022 \ac{asd} dataset explicitly features the \ac{dg} problem for \ac{asd}.
This means that $990$ normal training samples belonging to a source domain and only $10$ normal training samples belonging to a target domain are provided for each of the machines.
The machine-specific test splits each consist of $200$ samples that can belong to any of the two domains and may be normal or anomalous.
In contrast to the DCASE2020 \ac{asd} dataset, the duration of individual recordings is not fixed but ranges between \SI{6}{\second} and \SI{18}{\second}.
Moreover, additional meta information, referred to as attribute information, about specific machine settings or the acoustic environment are provided for each recording.
\par
\textbf{DCASE2023:} The DCASE2023 \ac{asd} dataset increases the difficulty of the \ac{asd} task with the following two modifications, which are jointly referred to by the term \enquote{first-shot problem}.
The first modification is that the development set and the evaluation set contain mutually exclusive machine types.
This ensures that \ac{asd} systems need to work well for arbitrary machine types as participants of the challenge cannot fine-tune their \ac{asd} systems to perform well for specific machine types.
The second modification is that, for some machine types, only recordings of a single machine are available.
This is more realistic but also substantially degrades the performance of discriminative embedding models as less information needs to be captured within the embeddings to obtain correct classification results.
\par
\textbf{DCASE2024:} Apart from exchanging some of the machine types contained in the dataset, the DCASE2024 \ac{asd} dataset has the following key differences compared to the DCASE2023 \ac{asd} dataset.
For some of the machine types, no attribute information is provided.
This means that the embedding model needs to be trained without utilizing any additional meta information for some machine types, which substantially simplifies the imposed auxiliary classification tasks used for training the models and leads to less informative embeddings.
Furthermore, some machines have an exclusive background noise, which means that embedding models trained to identify these machines may only be monitoring the noise.
As the background noise does not carry any relevant information for detecting anomalous machine sounds, this strongly degrades the performance when using such an embedding model.
\par
\textbf{DCASE2025:} The main modification of the DCASE 2025 \acs{asd} dataset is that the dataset also contains supplemental data for each machine type, which may either contain clean recordings of the target machines or only background noise.
However, to simplify the evaluations in this work, we restrained from specific adaptations of the \ac{asd} systems and simply ignored the supplemental data.
Thus, the main purpose of using the dataset in the experiments of this work is to have additional evaluations with recordings belonging to different machine types than the ones contained in previous versions of the dataset.

\subsection{Evaluation metrics}
To evaluate the performance of the \ac{asd} systems, we followed the official evaluation metrics for \ac{asd} experiments with each individual dataset \cite{koizumi2020description,dohi2023description,nishida2024description}.
For the DCASE2020 dataset, the \ac{auc} and \ac{pauc} \cite{mcclish1989analyzing} with $p=0.1$ are computed for each section, and then the arithmetic mean is taken over all results.
For the DCASE2023 and DCASE2024 datasets, the section-specific \acp{auc} are computed individually for each domain by considering only the normal test samples belonging to the source or target domain but using all anomalous test samples regardless of the domain they belong to.
The \acp{pauc} are computed using all test samples belonging to any domain.
Finally, the harmonic mean is taken over both \acp{auc} and the \ac{pauc} of all sections.

\subsection{Embedding models}
\label{sec:models}
For the experiments conducted in this work, four different embedding models were used.
The first one (Direct-\acs{act}) is trained with an auxiliary classification task based on meta information and \ac{ssl}.
The other three (OpenL3-raw, BEATs-raw, and \acs{eat}-raw) are pre-trained general-purpose models that are used without further training.
In addition, we used an ensemble of ten Direct-\acs{act} models to evaluate the normalization approach.
In the following, the models are described and their implementation details are provided.
\par
\textbf{Direct-\acs{act}:} This embedding model is directly trained on the data using an auxiliary classification task and does not depend on any pre-trained models.
More specifically, this model is based on \cite{wilkinghoff2024self} and consists of two feature branches.
The first branch utilizes the magnitude of the full spectrum, i.e., the Fourier transform of the entire signal.
The second branch computes the magnitude of the \ac{stft} with a Hann window of size $1024$ and a step size of $512$, and applies temporal mean normalization to remove constant frequency information and thus make both features more different from each other.
A different convolutional neural network is applied to each feature branch, mapping each of the features into a $256$-dimensional embedding space.
After that, both resulting embeddings are concatenated to obtain a single embedding.
The \ac{act} used to train the embedding model is defined based on the provided meta information, with one vanilla sub-task on the concatenated embeddings and an additional \ac{ssl} sub-task on augmented embeddings.
In the vanilla sub-task, the model has to discriminate, based on the concatenated feature embedding, between different values of the provided meta information, more specifically the machine types, machine IDs and additional parameter settings, or information about the acoustic environment.
The additional \ac{ssl} sub-task uses feature exchange \cite{wilkinghoff2024self}, which randomly exchanges the embeddings of the two feature branches between samples and asks the model to predict whether the embeddings belong to the same sample or not, on top of predicting the meta information associated with each sample.
For the entire dataset, a single embedding model is trained for $10$, $5$, and $15$ epochs on the DCASE2020, DCASE2023, and DCASE2024 dataset, respectively, using Adam \cite{kigma2015adam} with a batch size of $64$.
These particular hyperparameters were chosen by optimizing the performance on the corresponding development sets.
The loss function is the AdaProj loss \cite{wilkinghoff2024adaproj}, which projects the data into class-specific linear subspaces on the unit sphere and ensures an angular margin between the classes.
For data augmentation, mixup \cite{zhang2017mixup} with a uniformly distributed mixing coefficient is applied to the waveforms.
Note that the only differences of this \ac{asd} system from the original system presented in \cite{wilkinghoff2024self} are that statistics exchange  \cite{chen2023effective} is not used and that the sub-cluster AdaCos loss \cite{wilkinghoff2021sub} is replaced with the AdaProj loss \cite{wilkinghoff2024adaproj} with a sub-space dimension of $32$.
\par
\textbf{OpenL3-raw:}
To show that the proposed normalization scheme does not depend on the choice of the embeddings and can also be utilized for pre-trained embeddings, the following \ac{asd} system based on OpenL3 embeddings \cite{cramer2019look}, which is an open-source model for Look, Listen, and Learn embeddings \cite{arandjelovic2017look,arandjelovic2018objects}, is used for additional evaluations.
First, all waveforms are divided into chunks using a sliding window with a length of \SI{1}{\second} and a hop size of \SI{0.1}{\second}.
Then, OpenL3 embeddings with a dimension of $512$ are extracted from these chunks by computing mel spectrograms with $128$ mel bins and using the model pre-trained on the environmental sound subset.
Finally, the (temporal) mean over all embeddings belonging to the chunks is taken to obtain a single embedding for each recording that serves as an input feature for a task-specific embedding model.
To calculate anomaly scores based on these embeddings, the \ac{mse} is used instead of the cosine distance.
\par
\textbf{BEATs-raw:}
To provide more evidence for the claim that the presented normalization approach does not depend on the input features, additional experiments with BEATs embeddings \cite{chen2023beats} were conducted.
In recent works on \ac{asd} \cite{jiang2024anopatch,saengthong2024deep}, systems based on these embeddings performed remarkably well.
For the experiments conducted in this work, the official BEATs model pre-trained for three iterations on Audioset \cite{gemmeke2017audioset} without additional fine-tuning was used.
To obtain a vector-sized embedding for each recording that serves as an input feature, the temporal mean of the patch embeddings is flattened to preserve frequency and channel information as proposed in \cite{saengthong2024deep}.
This results in a single BEATs embedding with a dimension of $6144$ for each audio recording.
Again, the cosine distance is replaced with the \ac{mse} when computing anomaly scores.
We also tried to train \acs{act} models by replacing the mel spectrogram input with BEATs and openL3 embeddings, but found that just using the off-the-shelf embeddings resulted in superior performance, a result consistent with~\cite{saengthong2024deep}.
We also tried to fine-tune BEATs on the \ac{asd} task to replicate~\cite{jiang2024anopatch}, but again the raw embeddings provided the best performance.
\par
\textbf{\acs{eat}-raw:} Similar to BEATs, we also utilized \acs{eat} embeddings \cite{chen2024eat} as used by several other \ac{asd} systems \cite{fujimura2025asdkit,saengthong2025Sgenrep}.
More concretely, we used the official checkpoint of the large \acs{eat} model pre-trained for $20$ epochs on Audioset \cite{gemmeke2017audioset} without further fine-tuning.
To obtain vector-sized embeddings, the temporal mean of the patch embeddings is concatenated with the extracted CLS embedding, resulting in embeddings with a dimension of $6912$.
As with the other pre-trained models, the \ac{mse} is used to compute anomaly scores.
\par
\textbf{Direct-\ac{act} (Ensemble):}
As many state-of-the-art \ac{asd} systems are ensembles consisting of multiple models, we also utilized an ensemble for additional evaluations.
This ensemble model is obtained by averaging the resulting anomaly scores of ten independently trained direct-\ac{act} models with the architecture described above.

\section{Experimental results}
\label{sec:results}
\subsection{Comparison of normalization approaches}
\label{subsubsec:comparison}
\input{figs/comparison.tex}
\input{figs/norm_comp.tex}
As a first experiment, different \ac{dg} approaches are compared to our proposed normalization approach and a simple nearest neighbor baseline approach.
More concretely, we compared four alternatives to our proposed approach: 1) $K$-means with $K=16$ on the reference samples belonging to the source domain \cite{wilkinghoff2023design}, 2) \ac{smote} \cite{chawla2002smote} to balance the number of samples for both domains by randomly interpolating between $4$ reference samples of the target domain to synthesize additional target domain samples, 3) \ac{lof} \cite{breunig2000lof} to compute local density-based anomaly scores, and 4) a domain-specific standardization of test score distributions \cite{saengthong2024deep}.
All experiments are conducted on the DCASE2022, DCASE2023, DCAE2024, and DCASE2025 \ac{asd} datasets with all five \ac{asd} models, i.e., Direct-\acs{act}, OpenL3-raw, BEATs-raw, \acs{eat}-raw, and Direct-\acs{act} (Ensemble).
The results can be found in \cref{tab:embs} and the following observations can be made.
\par
First, it can be seen that the performance in the target domain is much worse than in the source domain when no \ac{dg} approach is applied, which verifies the motivation for this work.
Second, all approaches improve the performance in the target domain while reducing the performance in the source domain, although to different extents.
Applying $K$-means to the reference samples of the source domain leads to moderate performance gains for Direct-\ac{act} and the ensemble model, but actually reduces the performance for OpenL3-raw, BEATs-raw, and EAT-raw.
Among all \ac{dg} approaches, \ac{smote} achieves the best performance in the source domain but only leads to minor improvements in the target domain.
Still, consistent improvements can be observed.
\Ac{lof}, on the other hand, leads to the worst performance in the source domain while achieving significant performance gains in the target domain.
Overall, the performance of \ac{lof} in the mixed domain is similar to the one achieved with \ac{smote}.
A domain-wise standardization of the anomaly scores and the difference-based and ratio-based variants of our proposed anomaly score normalization approach lead to the highest performance gains in the target domain while not reducing the performance in the source domain too much.
As a result, these three approaches all achieve similar and highest overall performance with no clear winner in terms of pure performance.
As a minor observation, it can be seen that our normalization approach also increases the performance gains obtained with additive ensembles.
The most likely reason for this is that the anomaly scores are scaled more similarly after normalizing them.
\par
Apart from comparing the effectiveness of the considered approaches in terms of performance, we also investigated other advantages and disadvantages of individual approaches in \cref{tab:norm_comparison}.
From that, it can be seen that our approach offers several advantages.
In contrast to using 
%source 
$K$-means in the source domain, balancing with \ac{smote}, or using a domain-specific standardization of anomaly scores, our approach does not require domain labels for the reference samples.
This is a clear advantage, since precisely defining domains and obtaining labels is difficult in real-world applications.
Moreover, \ac{smote} and standardizing the score distributions are adaptations to specific domain shifts, which require modifying the system for every new domain shift that occurs. Last but not least, domain-specific standardization (as implemented in~\cite{saengthong2025Sgenrep}), which is the only approach resulting in similar performance improvements as our proposed approach, estimates first- and second-order statistics of the anomaly scores of all test samples to modify the scores. Therefore, test samples cannot be independently evaluated, which is a strong restriction for real-world applications.
We also tried to estimate these statistics using the anomaly scores of the training samples, but this only degraded the performance instead of improving it.
In summary, our proposed approach yields the highest performance while not suffering from any of the mentioned disadvantages, and thus is a favorable choice among all \ac{dg} approaches.

\subsection{Sensitivity analysis with respect to the hyperparameters}

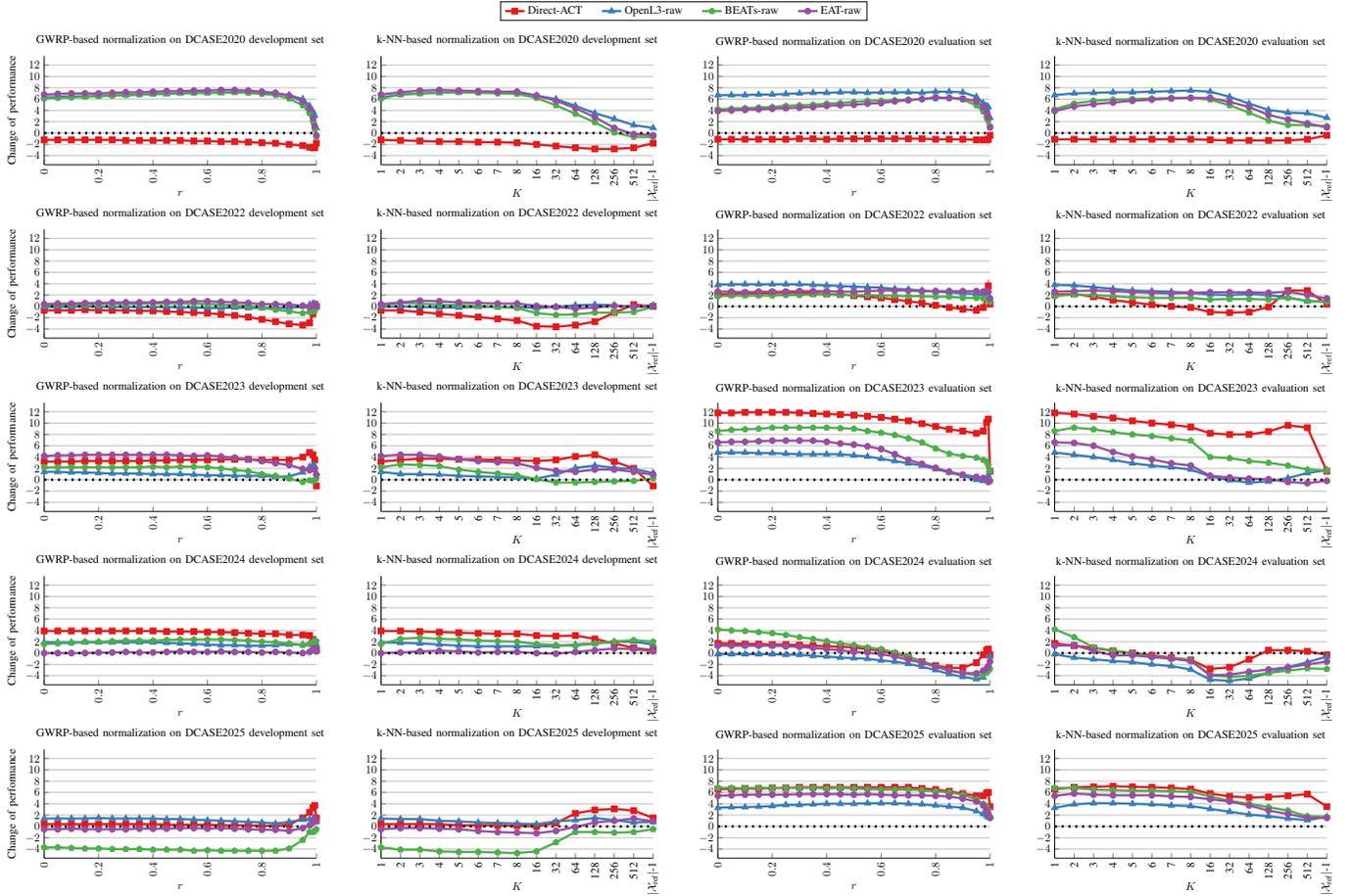
\begin{figure*}
    \centering
    \begin{adjustbox}{max width=\textwidth}
          \input{figs/scaling_ablation.tikz}
    \end{adjustbox}
    \caption{Performance change for the ratio-based score normalization using different values of the \acs{gwrp} constant $r$ and number $K$ for \acs{knn}. The models Direct-\ac{act}, OpenL3-raw, BEATs-raw, and \acs{eat}-raw are evaluated on the DCASE2020, DCASE2022, DCASE2023, DCASE2024, and DCASE2025 datasets. For Direct-\ac{act}, mean results over ten independent trials are shown. Similar trends can be seen in the plots corresponding to the difference-based normalization.}
    \label{fig:scaling_ablation}
\end{figure*}
As additional ablation studies, we investigated tuning the hyperparameters $r$ of \ac{gwrp} and $K$ of \ac{knn} for the respective parameterizations of the proposed normalization approach.
The performance changes on the DCASE2020, DCASE2022, DCASE2023, DCASE2024, and DCASE2025 datasets for the embedding models Direct-\acs{act}, OpenL3-raw, BEATs-raw and EAT-raw when normalizing the scores are depicted in \cref{fig:scaling_ablation}.
The following observations can be made.
\par
First and most importantly, the proposed approach improves the performance for most datasets if local density estimates are used. 
There are a few exceptions to this, for example the performance obtained with Direct-\ac{act} on the development set of the DCASE2020 and DCASE2022 datasets, which slightly degrades.
However, since the DCASE2020 dataset does not contain any domain shifts, the normalization approach does not need to be applied.
Interestingly, the performance gains for OpenL3-raw, BEATs-raw, and EAT-raw on this dataset are still substantial, which indicates that the proposed normalization approach is beneficial for off-the-shelf embeddings when not using any fine-tuning.
For the DCASE2022 dataset, the performance degradation with direct-\ac{act} can be explained by a slightly larger decrease in performance on the source domain compared to the improvement achieved on the target domain, resulting in a marginal overall decline.
Other exceptions are the performance obtained with OpenL3-raw on the evaluation set of the DCASE2024 dataset and with BEATs-raw on the development set of the DCASE2025 dataset, which significantly degrades, for an unknown reason.
\par
The second main observation is that optimal hyperparameter values are strongly dataset-dependent and partly embedding-dependent.
Although there are several cases where increasing the hyperparameter slightly improves the performance, e.g., for OpenL3-raw, BEATs-raw, and EAT-raw on the DCASE2020 dataset or for all embedding models on the DCASE2025 development set, these improvements are not consistent on all datasets.
This is particularly evident on the evaluation set of the DCASE2024 dataset, where performance drops rapidly when increasing the value of $K$ and $r$. In contrast, very small values lead to consistent improvements in performance. Thus, we recommend being conservative and only using a single sample to define the local density, i.e., $K=1$ or $r=0$ without additional prior knowledge. Note that these numbers are different from the results presented in \cite{wilkinghoff2024keeping}, as the results presented here are based on the official performance metric of the DCASE Challenge, while a simplified performance measure was used in \cite{wilkinghoff2024keeping}.

\subsection{Comparison to the state of the art}
\input{figs/sota.tex}
In \cref{tab:sota}, the proposed anomaly score normalization approach is evaluated on a representative group of \ac{asd} datasets, namely the DCASE2020, DCASE2023, and DCASE2024 datasets, and the resulting performance is compared to the state of the art.
For this comparison, we consider the five systems direct-\ac{act}, OpenL3-raw, BEATs-raw, EAT-raw, and direct-\ac{act} (ensemble) as described in \cref{sec:models}, all with a local-density-based normalization of the anomaly scores using \ac{knn} with $K=1$.
While the direct-\ac{act} model with the proposed normalization stays slightly behind the state-of-the-art performance, the ensemble robustly outperforms the state-of-the-art systems on the DCASE2020 and DCASE2023 datasets.
On the DCASE2024 dataset, our performance is better on the development set but still worse than the state of the art and comparable to the baseline system \cite{harada2023first} on the evaluation set.
The main reason for this poorer performance is that some machine types contained in the DCASE2024 dataset have specific noise conditions.
For these machine types, an \ac{act}-based system only needs to monitor the noise and thus completely fails in detecting subtle changes in target machine sounds.
Another reason is that, for some machine types, no attribute information is provided, which further degrades the performance.
Both imposed difficulties need to be addressed to be able to achieve state-of-the-art performance.
Additional evidence for this claim is that the performance obtained with the simple BEATs-raw model is only slightly worse than the direct-\ac{act} ensemble on the DCASE2024 dataset while being much worse on the DCASE2020 and DCASE2023 datasets.
Note that BEATs and AnoPatch make extensive use of \ac{ssl} via masking patches while the direct-\ac{act} model only uses \ac{ssl} via feature exchange, similarly to OpenL3. The difference in performance can thus be seen as evidence for the importance of using \ac{ssl} to learn suitable representations for \ac{asd}.

\section{Conclusions and future work}
\label{sec:conclusions}
In this work, a simple yet highly effective anomaly score normalization approach for \ac{dg} was presented.
The approach was extensively evaluated on the DCASE2020, DCASE2022, DCASE2023, DCASE2024, and DCASE2025 \ac{asd} datasets using state-of-the-art embedding models, from a model based on directly training with the discriminative angular margin loss AdaProj, to models based on pre-trained embeddings, such as OpenL3, BEATs, and EAT.
The proposed normalization approach was shown to consistently and significantly improve the performance in domain-shifted conditions and outperforms other existing anomaly score calculation and normalization approaches while not using any domain labels or adapting to specific target domains, and treating each test sample independently.
As a result, an ensemble-based \ac{asd} system was presented that utilizes this normalization approach and achieves state-of-the-art performance on the DCASE2020 and DCASE2023 dataset.
For future work, we plan to evaluate the proposed normalization approach for other applications and different data modalities in addition to audio.
Furthermore, we plan to explore synergies with other normalization approaches, as also done in \cite{saengthong2025Sgenrep}.
Last but not least, the proposed \ac{asd} system will be adapted to provide state-of-the-art performance on the DCASE2024 dataset by fine-tuning BEATs with an \ac{act}, similarly to AnoPatch \cite{jiang2024anopatch}.

\bibliographystyle{IEEEtran}
\bibliography{refs}

\iffalse
% the biography section is optional
\newpage

\section{Biography Section}
If you have an EPS/PDF photo (graphicx package needed), extra braces are
 needed around the contents of the optional argument to biography to prevent
 the LaTeX parser from getting confused when it sees the complicated
 $\backslash${\tt{includegraphics}} command within an optional argument. (You can create
 your own custom macro containing the $\backslash${\tt{includegraphics}} command to make things
 simpler here.)
 
\vspace{11pt}

\bf{If you include a photo:}\vspace{-33pt}
\begin{IEEEbiography}[{\includegraphics[width=1in,height=1.25in,clip,keepaspectratio]{fig1}}]{Michael Shell}
Use $\backslash${\tt{begin\{IEEEbiography\}}} and then for the 1st argument use $\backslash${\tt{includegraphics}} to declare and link the author photo.
Use the author name as the 3rd argument followed by the biography text.
\end{IEEEbiography}

\vspace{11pt}

\bf{If you will not include a photo:}\vspace{-33pt}
\begin{IEEEbiographynophoto}{John Doe}
Use $\backslash${\tt{begin\{IEEEbiographynophoto\}}} and the author name as the argument followed by the biography text.
\end{IEEEbiographynophoto}
\fi

\vfill

\end{document}

%% file: figs/datasets.tex
\begin{table*}[t]
	\centering
    \sisetup{
    detect-weight, % Make siunitx detect align bold cells correctly
    mode=text, % Make siuntix print tables in text mode (causes width of bold characters to be the same as non-bold)
    tight-spacing=true,
    round-mode=places,
    round-precision=0,
    table-format=3,
    table-number-alignment=center
}
	\caption{Overview of the considered DCASE \ac{asd} datasets.}%\vspace{-8pt}
\begin{adjustbox}{max width=\textwidth}
	\begin{tabular}{lS[table-format=2]ccccccccccc}
		\toprule
        &&&&&&&&&\multicolumn{4}{c}{\# recordings (per section)}\\
        \cmidrule(lr){10-13}
        &\multicolumn{3}{c}{\# machine types}&\multicolumn{3}{c}{\# sections (per machine type)}&&&\multicolumn{2}{c}{source domain}&\multicolumn{2}{c}{target domain}\\
        \cmidrule(lr){2-4}\cmidrule(lr){5-7}\cmidrule(lr){10-11}\cmidrule(lr){12-13}
        Name&{total}&dev.\ set&eval.\ set&total&dev.\ set&eval.\ set&\# \acs{act} classes&split&{normal}&{anomalous}&{normal}&{anomalous}\\
		\midrule
		\multirow{2}{*}{DCASE2020 \cite{koizumi2020description}} & {\multirow{2}{*}{6}} & \multirow{2}{*}{6} & \multirow{2}{*}{6} & \multirow{2}{*}{6-7} & \multirow{2}{*}{3-4} & \multirow{2}{*}{3} & \multirow{2}{*}{41} & train & $\leq1000$ & 0 & 0 & 0\\
		&&&&&&&& test & $\leq400$ & $\leq200$ & 0 & 0\\
		\midrule
		\multirow{2}{*}{DCASE2022 \cite{dohi2022description}} & {\multirow{2}{*}{7}} & \multirow{2}{*}{7} & \multirow{2}{*}{7} & \multirow{2}{*}{6} & \multirow{2}{*}{3} & \multirow{2}{*}{3} & \multirow{2}{*}{242} & train & 990 & 0 & 10 & 0\\
		&&&&&&&& test & 50 & 50 & 50 & 50\\
		\midrule
		\multirow{2}{*}{DCASE2023 \cite{dohi2023description}} & {\multirow{2}{*}{14}} & \multirow{2}{*}{7} & \multirow{2}{*}{7} & \multirow{2}{*}{1} & \multirow{2}{*}{1} & \multirow{2}{*}{1} & \multirow{2}{*}{167} & train & 990 & 0 & 10 & 0\\
		&&&&&&&& test & 50 & 50 & 50 & 50\\
        \midrule
		\multirow{2}{*}{DCASE2024 \cite{nishida2024description}} & {\multirow{2}{*}{16}} & \multirow{2}{*}{7} & \multirow{2}{*}{9} & \multirow{2}{*}{1} & \multirow{2}{*}{1} & \multirow{2}{*}{1} & \multirow{2}{*}{141} & train & 990 & 0 & 10 & 0\\
		&&&&&&&& test & 50 & 50 & 50 & 50\\
		\midrule
		\multirow{2}{*}{DCASE2025 \cite{nishida2025description}} & {\multirow{2}{*}{14}} & \multirow{2}{*}{7} & \multirow{2}{*}{7} & \multirow{2}{*}{1} & \multirow{2}{*}{1} & \multirow{2}{*}{1} & \multirow{2}{*}{172} & train & 990 & 0 & 10 & 0\\
		&&&&&&&& test & 50 & 50 & 50 & 50\\
		\bottomrule
	\end{tabular}
\end{adjustbox}
\label{tab:datasets}
\end{table*}

%% file: figs/comparison.tex
\begin{table*}[!t]
\centering
\caption{Harmonic means of all AUCs and pAUCs obtained with different anomaly score calculation approaches and embedding models. All included values are the harmonic means over the performance metrics of all development and evaluation sets of the DCASE2022, DCASE2023, DCASE2024, and DCASE2025 \acs{asd} datasets. For non-deterministic models, means over ten independent trials corresponding to ten trained embedding models are shown. To allow for better comparison, the same ten trained embedding models were used for all evaluations. To obtain means of the source domain as reference samples, K-means with $K=16$ was applied. For \acs{smote}, four neighbors were used to synthesize samples. For \acs{lof} and both variants of the proposed normalization approach, $K=1$ was used. Highest numbers in each column and block for each embedding model are in bold.}
\begin{adjustbox}{max width=0.99\textwidth}
\begin{NiceTabular}{c*{7}{c}}
\toprule
\ac{dg} approach&domain&Direct-\acs{act}&OpenL3-raw&BEATs-raw&EAT-raw&Direct-\acs{act} (ensemble)&average\\
\midrule
\multicolumn{1}{c}{-}&\multicolumn{1}{c}{source}
&\pmb{$66.0\%$} % hmean([hmean([74.4,71.1]),hmean([71.9,75.6]),hmean([65.4,60.4]),hmean([60.2,55.0])])
&\pmb{$62.6\%$} % hmean([hmean([63.8,60.6]),hmean([62.3,68.4]),hmean([65.6,62.2]),hmean([58.5,60.7])])
&\pmb{$65.0\%$} % hmean([hmean([67.9,64.2]),hmean([65.2,74.6]),hmean([61.8,61.3]),hmean([63.2,63.5])])
&\pmb{$64.0\%$} % hmean([hmean([65.9,63.3]),hmean([63.4,71.2]),hmean([60.3,62.9]),hmean([62.8,62.9])])
&\pmb{$67.2\%$} % hmean([hmean([75.4,72.4]),hmean([73.1,78.0]),hmean([67.1,61.7]),hmean([60.7,55.9])])
&\pmb{$65.0\%$}\\
\multicolumn{1}{c}{using source means \cite{wilkinghoff2023design}}&\multicolumn{1}{c}{source}
&$62.4\%$ % hmean([hmean([73.8,71.9]),hmean([68.8,70.3]),hmean([61.8,52.6]),hmean([57.3,51.6])])
&$59.8\%$ % hmean([hmean([62.8,59.3]),hmean([61.8,64.5]),hmean([56.8,58.6]),hmean([57.7,57.8])])
&$63.2\%$ % hmean([hmean([66.8,62.3]),hmean([64.1,69.0]),hmean([59.0,59.9]),hmean([63.6,62.3])])
&$61.4\%$ % hmean([hmean([65.4,61.6]),hmean([60.1,64.9]),hmean([58.7,58.6]),hmean([62.1,60.9])])
&$63.6\%$ % hmean([hmean([75.1,73.5]),hmean([70.0,73.4]),hmean([63.4,53.5]),hmean([57.9,51.8])])
&$62.1\%$\\
\multicolumn{1}{c}{\acs{smote} \cite{chawla2002smote}}&\multicolumn{1}{c}{source}
&$65.6\%$ % hmean([hmean([74.2,72.3]),hmean([71.7,75.3]),hmean([64.8,58.9]),hmean([59.7,54.9])])
&$60.6\%$ % hmean([hmean([63.4,60.4]),hmean([62.8,66.7]),hmean([57.8,57.8]),hmean([57.9,59.4])])
&$63.7\%$ % hmean([hmean([66.9,64.3]),hmean([64.7,73.2]),hmean([61.8,57.8]),hmean([61.9,61.4])])
&$63.1\%$ % hmean([hmean([65.8,63.1]),hmean([63.1,70.2]),hmean([60.1,60.9]),hmean([61.8,60.9])])
&$66.7\%$ % hmean([hmean([75.3,73.6]),hmean([73.0,77.7]),hmean([66.3,60.3]),hmean([60.1,54.8])])
&$63.9\%$\\
\multicolumn{1}{c}{\acs{lof} \cite{breunig2000lof}}&\multicolumn{1}{c}{source}
&$58.9\%$ % hmean([hmean([64.9,63.3]),hmean([61.3,67.4]),hmean([57.3,51.0]),hmean([55.7,54.2])])
&$58.1\%$ % hmean([hmean([59.2,60.9]),hmean([57.7,66.0]),hmean([56.7,54.0]),hmean([57.3,54.7])])
&$60.0\%$ % hmean([hmean([59.7,61.5]),hmean([59.8,67.0]),hmean([58.9,54.9]),hmean([59.0,60.3])])
&$58.2\%$ % hmean([hmean([58.5,58.2]),hmean([59.6,62.7]),hmean([56.5,54.2]),hmean([57.8,59.2])])
&$62.1\%$ % hmean([hmean([69.9,67.9]),hmean([65.0,73.0]),hmean([60.8,52.1]),hmean([58.5,55.8])])
&$59.5\%$\\
\multicolumn{1}{c}{standardization \cite{saengthong2024deep}}&\multicolumn{1}{c}{source}
&$63.6\%$ % hmean([hmean([74.3,71.1]),hmean([70.3,71.8]),hmean([61.5,56.2]),hmean([57.4,53.4])])
&$59.0\%$ % hmean([hmean([62.3,59.3]),hmean([61.7,63.8]),hmean([56.4,58.1]),hmean([55.5,56.0])])
&$61.5\%$ % hmean([hmean([66.4,62.3]),hmean([62.5,69.7]),hmean([58.4,58.0]),hmean([58.4,58.3])])
&$60.2\%$ % hmean([hmean([64.3,60.8]),hmean([60.0,66.2]),hmean([56.8,59.1]),hmean([58.3,57.3])])
&$65.0\%$ % hmean([hmean([75.7,72.8]),hmean([72.0,75.2]),hmean([63.1,57.4]),hmean([57.9,54.2])])
&$61.9\%$\\
\multicolumn{1}{c}{normalization (difference)}&\multicolumn{1}{c}{source}
&$62.3\%$ % hmean([hmean([71.0,69.2]),hmean([68.2,69.2]),hmean([62.8,53.6]),hmean([55.4,55.4])])
&$59.3\%$ % hmean([hmean([60.9,62.4]),hmean([59.5,66.2]),hmean([57.3,52.9]),hmean([58.4,58.6])])
&$63.4\%$ % hmean([hmean([65.9,63.8]),hmean([64.6,70.6]),hmean([61.9,60.2]),hmean([57.9,63.8])])
&$62.2\%$ % hmean([hmean([63.7,62.7]),hmean([65.1,68.6]),hmean([59.4,57.2]),hmean([60.5,61.9])])
&$64.8\%$ % hmean([hmean([73.3,72.2]),hmean([70.6,72.9]),hmean([65.6,55.1]),hmean([58.2,57.1])])
&$62.4\%$\\
\multicolumn{1}{c}{normalization (ratio)}&\multicolumn{1}{c}{source}
&$60.6\%$ % hmean([hmean([67.7,65.5]),hmean([65.7,70.0]),hmean([61.8,50.5]),hmean([53.9,55.7])])
&$57.8\%$ % hmean([hmean([59.7,61.3]),hmean([56.4,66.4]),hmean([52.3,52.3]),hmean([57.9,59.0])])
&$62.0\%$ % hmean([hmean([64.1,61.4]),hmean([62.1,70.1]),hmean([61.3,60.0]),hmean([55.2,63.8])])
&$61.6\%$ % hmean([hmean([63.1,61.6]),hmean([64.7,68.4]),hmean([59.4,56.1]),hmean([58.9,61.9])])
&$63.1\%$ % hmean([hmean([70.7,68.3]),hmean([69.6,75.4]),hmean([65.1,51.3]),hmean([55.0,58.0])])
&$61.0\%$\\
\midrule
\multicolumn{1}{c}{-}&\multicolumn{1}{c}{target}
&$52.5\%$ % hmean([hmean([64.0,57.8]),hmean([59.1,48.0]),hmean([52.4,47.2]),hmean([51.1,45.9])])
&$51.7\%$ % hmean([hmean([54.3,54.9]),hmean([54.5,52.2]),hmean([49.2,49.8]),hmean([50.3,49.4])])
&$54.2\%$ % hmean([hmean([56.7,56.7]),hmean([57.9,51.1]),hmean([52.0,53.7]),hmean([57.2,49.6])])
&$53.1\%$ % hmean([hmean([55.2,54.8]),hmean([55.9,52.4]),hmean([52.7,52.0]),hmean([55.8,47.2])])
&$52.6\%$ % hmean([hmean([65.2,58.2]),hmean([59.3,47.3]),hmean([53.0,47.0]),hmean([51.7,45.3])])
&$52.8\%$\\
\multicolumn{1}{c}{using source means \cite{wilkinghoff2023design}}&\multicolumn{1}{c}{target}
&$58.7\%$ % hmean([hmean([65.7,60.8]),hmean([63.0,57.3]),hmean([58.4,54.0]),hmean([54.8,57.4])])
&$52.3\%$ % hmean([hmean([55.2,55.1]),hmean([55.3,52.7]),hmean([51.5,49.6]),hmean([50.7,48.9])])
&$53.4\%$ % hmean([hmean([57.2,57.6]),hmean([57.5,48.5]),hmean([51.4,52.0]),hmean([57.1,48.1])])
&$53.8\%$ % hmean([hmean([55.6,55.1]),hmean([55.3,52.3]),hmean([53.1,52.9]),hmean([57.2,49.6])])
&$59.6\%$ % hmean([hmean([66.5,61.6]),hmean([63.5,59.0]),hmean([59.4,54.6]),hmean([55.1,59.1])])
&$55.6\%$\\
\multicolumn{1}{c}{\acs{smote} \cite{chawla2002smote}}&\multicolumn{1}{c}{target}
&$55.6\%$ % hmean([hmean([65.0,59.7]),hmean([61.6,52.9]),hmean([55.6,50.0]),hmean([52.5,50.8])])
&$55.0\%$ % hmean([hmean([54.9,56.5]),hmean([56.2,57.1]),hmean([53.1,56.7]),hmean([51.7,54.4])])
&$57.5\%$ % hmean([hmean([58.1,58.0]),hmean([59.7,58.9]),hmean([53.0,58.5]),hmean([59.6,54.6])])
&$55.7\%$ % hmean([hmean([56.2,55.9]),hmean([57.2,57.9]),hmean([53.5,56.4]),hmean([57.8,51.6])])
&$56.0\%$ % hmean([hmean([66.0,60.4]),hmean([62.2,53.0]),hmean([56.2,50.0]),hmean([52.8,51.2])])
&$56.0\%$\\
\multicolumn{1}{c}{\acs{lof} \cite{breunig2000lof}}&\multicolumn{1}{c}{target}
&$59.5\%$ % hmean([hmean([63.8,63.5]),hmean([63.0,62.2]),hmean([57.3,57.5]),hmean([57.7,52.6])])
&$56.5\%$ % hmean([hmean([56.9,58.8]),hmean([61.6,57.6]),hmean([53.3,58.8]),hmean([53.1,52.9])])
&$58.4\%$ % hmean([hmean([58.3,59.8]),hmean([62.8,61.2]),hmean([52.8,61.3]),hmean([57.5,55.1])])
&$56.6\%$ % hmean([hmean([56.8,57.8]),hmean([61.7,58.6]),hmean([50.8,60.7]),hmean([56.1,51.9])])
&$62.2\%$ % hmean([hmean([67.1,66.2]),hmean([66.1,66.6]),hmean([60.2,59.8]),hmean([60.2,53.8])])
&$58.6\%$\\
\multicolumn{1}{c}{standardization \cite{saengthong2024deep}}&\multicolumn{1}{c}{target}
&$60.0\%$ % hmean([hmean([65.9,63.6]),hmean([64.9,63.1]),hmean([59.1,54.9]),hmean([55.6,55.5])])
&$56.2\%$ % hmean([hmean([55.9,57.5]),hmean([57.0,59.6]),hmean([54.4,55.7]),hmean([53.1,56.9])])
&$59.4\%$ % hmean([hmean([58.3,59.1]),hmean([59.8,63.5]),hmean([55.3,59.6]),hmean([60.2,60.3])])
&$57.6\%$ % hmean([hmean([56.9,56.9]),hmean([56.7,62.3]),hmean([54.6,58.7]),hmean([59.0,56.6])])
&$61.2\%$ % hmean([hmean([67.1,64.5]),hmean([66.2,64.9]),hmean([60.4,55.9]),hmean([56.2,56.7])])
&$58.9\%$\\
\multicolumn{1}{c}{normalization (difference)}&\multicolumn{1}{c}{target}
&$60.9\%$ % hmean([hmean([67.1,65.5]),hmean([66.0,62.8]),hmean([59.6,57.3]),hmean([57.4,54.3])])
&$57.5\%$ % hmean([hmean([57.1,59.9]),hmean([61.1,59.3]),hmean([55.0,58.1]),hmean([54.9,55.0])])
&\pmb{$61.2\%$} % hmean([hmean([60.5,62.2]),hmean([65.3,62.9]),hmean([56.3,63.2]),hmean([61.4,58.7])])
&$59.3\%$ % hmean([hmean([59.3,61.0]),hmean([62.4,63.0]),hmean([53.6,60.9]),hmean([60.1,55.6])])
&$63.4\%$ % hmean([hmean([69.5,68.0]),hmean([67.4,66.3]),hmean([61.7,59.8]),hmean([58.9,57.7])])
&$60.5\%$\\
\multicolumn{1}{c}{normalization (ratio)}&\multicolumn{1}{c}{target}
&\pmb{$61.5\%$} % hmean([hmean([66.7,65.0]),hmean([67.8,63.8]),hmean([60.4,58.4]),hmean([57.7,54.7])])
&\pmb{$58.0\%$} % hmean([hmean([57.4,59.7]),hmean([62.4,59.3]),hmean([58.3,58.3]),hmean([54.6,55.0])])
&\pmb{$61.2\%$} % hmean([hmean([60.7,61.8]),hmean([66.2,62.6]),hmean([55.4,63.8]),hmean([61.2,59.0])])
&\pmb{$59.4\%$} % hmean([hmean([59.4,60.2]),hmean([64.1,62.4]),hmean([53.9,60.2]),hmean([60.3,56.1])])
&\pmb{$64.1\%$} % hmean([hmean([69.5,67.9]),hmean([70.9,67.3]),hmean([63.1,61.2]),hmean([59.7,56.2])])
&\pmb{$60.8\%$}\\
\midrule
\multicolumn{1}{c}{-}&\multicolumn{1}{c}{mixed}
&$58.0\%$ % hmean([hmean([68.4,64.0]),hmean([65.2,56.2]),hmean([58.1,53.0]),hmean([55.7,49.0])])
&$56.7\%$ % hmean([hmean([59.4,57.7]),hmean([59.1,59.0]),hmean([54.7,55.9]),hmean([54.6,54.0])])
&$59.3\%$ % hmean([hmean([62.2,60.5]),hmean([62.5,59.0]),hmean([56.5,58.2]),hmean([61.4,55.2])])
&$58.5\%$ % hmean([hmean([60.8,59.1]),hmean([60.8,59.5]),hmean([56.8,57.6]),hmean([60.3,53.5])])
&$58.3\%$ % hmean([hmean([69.4,64.6]),hmean([65.9,55.7]),hmean([58.8,53.6]),hmean([56.2,48.1])])
&$58.2\%$\\
\multicolumn{1}{c}{using source means \cite{wilkinghoff2023design}}&\multicolumn{1}{c}{mixed}
&$61.1\%$ % hmean([hmean([70.0,66.3]),hmean([66.8,63.4]),hmean([61.3,53.9]),hmean([56.9,54.7])])
&$56.3\%$ % hmean([hmean([59.7,57.6]),hmean([59.6,58.1]),hmean([54.4,54.0]),hmean([54.8,52.9])])
&$58.2\%$ % hmean([hmean([62.4,60.4]),hmean([62.2,55.5]),hmean([55.6,56.1]),hmean([61.0,53.7])])
&$58.1\%$ % hmean([hmean([60.8,58.8]),hmean([60.1,58.1]),hmean([56.4,56.2]),hmean([60.6,54.7])])
&$62.3\%$ % hmean([hmean([71.2,67.5]),hmean([67.9,65.3]),hmean([62.8,55.0]),hmean([57.3,55.6])])
&$59.2\%$\\
\multicolumn{1}{c}{\acs{smote} \cite{chawla2002smote}}&\multicolumn{1}{c}{mixed}
&$60.3\%$ % hmean([hmean([69.2,65.7]),hmean([67.1,60.7]),hmean([60.4,54.6]),hmean([56.5,52.6])])
&$57.9\%$ % hmean([hmean([59.9,58.8]),hmean([60.5,62.6]),hmean([56.2,53.8]),hmean([55.4,57.2])])
&$61.1\%$ % hmean([hmean([62.8,61.4]),hmean([63.2,65.3]),hmean([57.6,59.4]),hmean([62.0,57.9])])
&$60.0\%$ % hmean([hmean([61.5,60.0]),hmean([61.4,64.1]),hmean([57.4,59.6]),hmean([61.0,55.8])])
&$60.9\%$ % hmean([hmean([70.1,66.5]),hmean([67.9,61.0]),hmean([61.2,55.5]),hmean([56.9,52.7])])
&$60.0\%$\\
\multicolumn{1}{c}{\acs{lof} \cite{breunig2000lof}}&\multicolumn{1}{c}{mixed}
&$59.9\%$ % hmean([hmean([65.2,64.0]),hmean([63.8,65.3]),hmean([58.1,54.3]),hmean([57.3,53.7])])
&$57.7\%$ % hmean([hmean([58.3,60.4]),hmean([60.6,62.1]),hmean([55.2,56.2]),hmean([55.6,54.4])])
&$59.8\%$ % hmean([hmean([59.1,61.3]),hmean([62.5,65.2]),hmean([56.0,57.9]),hmean([59.0,58.2])])
&$58.0\%$ % hmean([hmean([58.1,58.8]),hmean([61.9,62.3]),hmean([53.8,57.4]),hmean([57.5,55.2])])
&$62.9\%$ % hmean([hmean([69.2,67.3]),hmean([67.5,70.0]),hmean([61.6,56.1]),hmean([60.1,55.2])])
&$59.7\%$\\
\multicolumn{1}{c}{standardization \cite{saengthong2024deep}}&\multicolumn{1}{c}{mixed}
&\pmb{$62.7\%$} % hmean([hmean([70.6,67.9]),hmean([68.8,68.4]),hmean([61.6,56.8]),hmean([57.3,54.7])])
&$58.5\%$ % hmean([hmean([60.0,59.1]),hmean([60.7,62.8]),hmean([56.1,57.9]),hmean([55.1,57.2])])
&$61.5\%$ % hmean([hmean([63.1,61.5]),hmean([62.4,67.6]),hmean([57.8,60.3]),hmean([60.6,59.7])])
&$60.0\%$ % hmean([hmean([61.5,59.7]),hmean([59.6,65.6]),hmean([56.5,60.4]),hmean([60.0,57.7])])
&$64.1\%$ % hmean([hmean([72.1,69.1]),hmean([70.0,71.2]),hmean([63.1,58.6]),hmean([57.8,55.7])])
&$61.4\%$\\
\multicolumn{1}{c}{normalization (difference)}&\multicolumn{1}{c}{mixed}
&$62.3\%$ % hmean([hmean([69.4,67.9]),hmean([68.1,66.5]),hmean([61.9,56.2]),hmean([56.9,55.2])])
&\pmb{$59.1\%$} % hmean([hmean([59.5,61.9]),hmean([62.0,63.5]),hmean([56.7,56.0]),hmean([56.6,57.3])])
&\pmb{$62.8\%$} % hmean([hmean([63.3,63.4]),hmean([66.3,67.4]),hmean([58.7,62.7]),hmean([59.6,61.7])])
&\pmb{$61.3\%$} % hmean([hmean([61.6,62.5]),hmean([64.8,66.4]),hmean([56.7,59.8]),hmean([60.9,58.7])])
&\pmb{$64.5\%$} % hmean([hmean([71.7,70.4]),hmean([70.0,70.5]),hmean([64.6,58.4]),hmean([58.2,56.8])])
&\pmb{$62.0\%$}\\
\multicolumn{1}{c}{normalization (ratio)}&\multicolumn{1}{c}{mixed}
&$61.8\%$ % hmean([hmean([67.7,66.1]),hmean([68.4,68.0]),hmean([62.0,54.7]),hmean([56.1,55.6])])
&$58.6\%$ % hmean([hmean([59.3,61.5]),hmean([60.5,63.8]),hmean([55.7,55.7]),hmean([56.0,57.3])])
&$62.0\%$ % hmean([hmean([62.4,62.3]),hmean([64.8,67.6]),hmean([58.1,62.4]),hmean([57.7,62.0])])
&$60.9\%$ % hmean([hmean([61.2,61.7]),hmean([65.0,66.1]),hmean([56.8,58.9]),hmean([59.8,58.9])])
&$64.3\%$ % hmean([hmean([70.3,68.6]),hmean([71.3,72.4]),hmean([65.2,56.5]),hmean([57.6,57.6])])
&$61.5\%$\\
\bottomrule
\end{NiceTabular}
\end{adjustbox}
\label{tab:embs}
\end{table*}

%% file: figs/norm_comp.tex
\begin{table*}
	\centering
	\caption{Qualitative comparison of different \ac{dg} approaches. Apart from the effectiveness in terms of performance, it is shown whether domain labels of the reference samples are required, whether the approach only adapts to specific target domains, and whether each test sample is evaluated independently from others. The performance provided for the effectiveness is the average mixed-domain performances over all datasets and embedding models as contained in \cref{tab:embs}.}%\vspace{-8pt}
\begin{adjustbox}{max width=\textwidth}
	\begin{tabular}{cccccc}
    % \cellcolor{green!30!white} very high
    % \cellcolor{green!15!white} high
    % \cellcolor{yellow!30!white} medium
    % \cellcolor{red!20} low
		\toprule
        \ac{dg} approach & requires domain labels & domain shift-specific & independent test samples & effectiveness (performance)\\
		\midrule
        - & \cellcolor{green!30!white} no & \cellcolor{green!30!white} no  & \cellcolor{green!30!white} yes& \cellcolor{red!20!white} low ($58.2\%$)\\
        using source means \cite{wilkinghoff2023design} & \cellcolor{red!20} yes & \cellcolor{green!30!white} no  & \cellcolor{green!30!white} yes& \cellcolor{orange!30!white} low/medium ($59.2\%$)\\
        SMOTE \cite{chawla2002smote} & \cellcolor{red!20} yes & \cellcolor{red!20} yes & \cellcolor{green!30} yes & \cellcolor{yellow!30!white} medium ($60.0\%$)\\
        \acs{lof} \cite{breunig2000lof} & \cellcolor{green!30!white} no & \cellcolor{green!30!white} no  & \cellcolor{green!30!white} yes& \cellcolor{yellow!30!white} medium ($59.7\%$)\\
        standardization \cite{saengthong2024deep} & \cellcolor{red!20} yes & \cellcolor{red!20} yes & \cellcolor{red!20} no & \cellcolor{green!15!white} high ($61.4\%$)\\
        proposed approach (difference) & \cellcolor{green!30!white} no & \cellcolor{green!30!white} no  & \cellcolor{green!30!white} yes & \cellcolor{green!15!white} high ($62.0\%$)\\
        proposed approach (ratio) & \cellcolor{green!30!white} no & \cellcolor{green!30!white} no  & \cellcolor{green!30!white} yes & \cellcolor{green!15!white} high ($61.5\%$)\\
		\bottomrule
	\end{tabular}
\end{adjustbox}
\label{tab:norm_comparison}
\end{table*}

%% file: figs/scaling_ablation.tikz
\begin{tikzpicture}
\begin{groupplot}[
    group style={
    group size=4 by 5,
    xlabels at=edge bottom,
    ylabels at=edge left,
    horizontal sep=2cm,vertical sep=2cm,},
	axis y line*=left,
    axis x line*=bottom,
    ymin=-4,
    ymax=12,
    enlarge y limits=0.1,
    legend style={at={(2.35,1.5)},anchor=north,legend columns=4,/tikz/every even column/.append style={column sep=0.5cm}},
    ylabel=Change of performance,
    height=4.95cm,
    width=10cm,
    ytick={-4,-2,0,2,4,6,8,10,12},
    xticklabel style={align=center,rotate=90},
    yticklabel style={align=center},
    xlabel near ticks,
    ylabel near ticks,
    ymajorgrids,
    cycle list/Set1
]
\nextgroupplot[title={\acs{gwrp}-based normalization on DCASE2020 development set},xlabel=$r$,
    xmin=0.0,
    xmax=1.0,]
\legend{Direct-\ac{act}, OpenL3-raw, BEATs-raw, EAT-raw}
% Direct-ACT
\addplot+[mark=square*,line width=2pt] coordinates {
(0,90.7-91.9)(0.05,90.7-91.9)(0.1,90.7-91.9)(0.15,90.7-91.9)(0.2,90.7-91.9)(0.25,90.7-91.9)(0.3,90.6-91.9)(0.35,90.6-91.9)(0.4,90.6-91.9)(0.45,90.6-91.9)(0.5,90.6-91.9)(0.55,90.5-91.9)(0.6,90.5-91.9)(0.65,90.4-91.9)(0.7,90.4-91.9)(0.75,90.3-91.9)(0.8,90.2-91.9)(0.85,90.1-91.9)(0.9,89.9-91.9)(0.95,89.7-91.9)(0.975,89.4-91.9)(0.9875,89.3-91.9)(0.99375,89.3-91.9)(1,90.1-91.9)};

% OpenL3-raw
\addplot+[mark=triangle*,line width=2pt] coordinates {
(0,75.2-68.9)(0.05,75.3-68.9)(0.1,75.4-68.9)(0.15,75.4-68.9)(0.2,75.5-68.9)(0.25,75.6-68.9)(0.3,75.6-68.9)(0.35,75.7-68.9)(0.4,75.8-68.9)(0.45,75.8-68.9)(0.5,75.9-68.9)(0.55,76.0-68.9)(0.6,76.0-68.9)(0.65,76.0-68.9)(0.7,76.1-68.9)(0.75,76.1-68.9)(0.8,76.1-68.9)(0.85,75.9-68.9)(0.9,75.6-68.9)(0.95,74.9-68.9)(0.975,73.7-68.9)(0.9875,72.7-68.9)(0.99375,72.0-68.9)(1,69.8-68.9)};

% BEATs-raw
\addplot+[mark=pentagon*,line width=2pt] coordinates {
(0,81.5-75.4)(0.05,81.6-75.4)(0.1,81.7-75.4)(0.15,81.8-75.4)(0.2,81.9-75.4)(0.25,82.0-75.4)(0.3,82.1-75.4)(0.35,82.2-75.4)(0.4,82.3-75.4)(0.45,82.4-75.4)(0.5,82.5-75.4)(0.55,82.5-75.4)(0.6,82.5-75.4)(0.65,82.6-75.4)(0.7,82.6-75.4)(0.75,82.5-75.4)(0.8,82.3-75.4)(0.85,82.1-75.4)(0.9,81.5-75.4)(0.95,80.3-75.4)(0.975,78.9-75.4)(0.9875,77.4-75.4)(0.99375,76.1-75.4)(1,74.7-75.4)};

% EAT-raw
\addplot+[mark=*,line width=2pt] coordinates {
(0,79.3-72.5)(0.05,79.4-72.5)(0.1,79.5-72.5)(0.15,79.5-72.5)(0.2,79.5-72.5)(0.25,79.6-72.5)(0.3,79.7-72.5)(0.35,79.7-72.5)(0.4,79.8-72.5)(0.45,79.9-72.5)(0.5,79.9-72.5)(0.55,80.0-72.5)(0.6,80.0-72.5)(0.65,80.1-72.5)(0.7,80.1-72.5)(0.75,80.0-72.5)(0.8,79.8-72.5)(0.85,79.6-72.5)(0.9,79.1-72.5)(0.95,78.1-72.5)(0.975,76.8-72.5)(0.9875,75.2-72.5)(0.99375,73.7-72.5)(1,72.1-72.5)};

\addplot[mark=none,line width=2pt,loosely dotted] coordinates {(0,0)(1,0)};

\nextgroupplot[title={\acs{knn}-based normalization on DCASE2020 development set},xlabel=$K$,xtick={1,2,3,4,5,6,7,8,9,10,11,12,13,14,15},xticklabels={1,2,3,4,5,6,7,8,16,32,64,128,256,512,$\lvert\cX_\text{ref}\rvert$-1},xmin=1,xmax=15,xlabel style={yshift=4ex}]
% Direct-ACT
\addplot+[mark=square*,line width=2pt] coordinates {
(1,90.7-91.9)(2,90.6-91.9)(3,90.5-91.9)(4,90.4-91.9)(5,90.4-91.9)(6,90.3-91.9)(7,90.3-91.9)(8,90.2-91.9)(9,89.9-91.9)(10,89.6-91.9)(11,89.3-91.9)(12,89.1-91.9)(13,89.1-91.9)(14,89.3-91.9)(15,90.1-91.9)};

% OpenL3-raw
\addplot+[mark=triangle*,line width=2pt] coordinates {
(1,75.2-68.9)(2,75.7-68.9)(3,75.9-68.9)(4,76.0-68.9)(5,76.1-68.9)(6,76.2-68.9)(7,76.1-68.9)(8,76.1-68.9)(9,75.5-68.9)(10,74.9-68.9)(11,73.7-68.9)(12,72.4-68.9)(13,71.4-68.9)(14,70.3-68.9)(15,69.8-68.9)};

% BEATs-raw
\addplot+[mark=pentagon*,line width=2pt] coordinates {
(1,81.5-75.4)(2,82.2-75.4)(3,82.4-75.4)(4,82.6-75.4)(5,82.6-75.4)(6,82.5-75.4)(7,82.4-75.4)(8,82.3-75.4)(9,81.6-75.4)(10,80.3-75.4)(11,78.8-75.4)(12,77.3-75.4)(13,75.5-75.4)(14,74.7-75.4)(15,74.7-75.4)};

% EAT-raw
\addplot+[mark=*,line width=2pt] coordinates {
(1,79.3-72.5)(2,79.7-72.5)(3,80.0-72.5)(4,80.1-72.5)(5,80.0-72.5)(6,79.9-72.5)(7,79.8-72.5)(8,79.8-72.5)(9,79.1-72.5)(10,78.2-72.5)(11,76.9-72.5)(12,75.3-72.5)(13,73.4-72.5)(14,72.3-72.5)(15,72.1-72.5)};

\addplot[mark=none,line width=2pt,loosely dotted] coordinates {(1,0)(15,0)};

\nextgroupplot[title={\acs{gwrp}-based normalization on DCASE2020 evaluation set},xlabel=$r$,
    xmin=0.0,
    xmax=1.0,]
% Direct-ACT
\addplot+[mark=square*,line width=2pt] coordinates {
(0,90.2-91.3)(0.05,90.2-91.3)(0.1,90.2-91.3)(0.15,90.2-91.3)(0.2,90.2-91.3)(0.25,90.2-91.3)(0.3,90.3-91.3)(0.35,90.3-91.3)(0.4,90.2-91.3)(0.45,90.3-91.3)(0.5,90.3-91.3)(0.55,90.3-91.3)(0.6,90.3-91.3)(0.65,90.3-91.3)(0.7,90.3-91.3)(0.75,90.3-91.3)(0.8,90.2-91.3)(0.85,90.2-91.3)(0.9,90.2-91.3)(0.95,90.1-91.3)(0.975,90.1-91.3)(0.9875,90.1-91.3)(0.99375,90.2-91.3)(1,90.9-91.3)};

% OpenL3-raw
\addplot+[mark=triangle*,line width=2pt] coordinates {
(0,77.5-70.8)(0.05,77.5-70.8)(0.1,77.5-70.8)(0.15,77.6-70.8)(0.2,77.6-70.8)(0.25,77.7-70.8)(0.3,77.8-70.8)(0.35,77.9-70.8)(0.4,77.9-70.8)(0.45,78.0-70.8)(0.5,78.0-70.8)(0.55,77.9-70.8)(0.6,78.0-70.8)(0.65,78.0-70.8)(0.7,78.0-70.8)(0.75,77.9-70.8)(0.8,78.1-70.8)(0.85,78.1-70.8)(0.9,78.0-70.8)(0.95,77.2-70.8)(0.975,76.1-70.8)(0.9875,75.6-70.8)(0.99375,75.4-70.8)(1,73.5-70.8)};

% BEATs-raw
\addplot+[mark=pentagon*,line width=2pt] coordinates {
(0,82.2-78.0)(0.05,82.3-78.0)(0.1,82.4-78.0)(0.15,82.5-78.0)(0.2,82.6-78.0)(0.25,82.8-78.0)(0.3,82.9-78.0)(0.35,83.0-78.0)(0.4,83.2-78.0)(0.45,83.3-78.0)(0.5,83.5-78.0)(0.55,83.7-78.0)(0.6,83.7-78.0)(0.65,83.8-78.0)(0.7,83.9-78.0)(0.75,84.0-78.0)(0.8,84.1-78.0)(0.85,84.2-78.0)(0.9,83.9-78.0)(0.95,82.9-78.0)(0.975,81.8-78.0)(0.9875,80.7-78.0)(0.99375,80.1-78.0)(1,79.2-78.0)};

% EAT-raw
\addplot+[mark=*,line width=2pt] coordinates {
(0,79.9-76.0)(0.05,80.0-76.0)(0.1,80.1-76.0)(0.15,80.2-76.0)(0.2,80.3-76.0)(0.25,80.4-76.0)(0.3,80.5-76.0)(0.35,80.6-76.0)(0.4,80.8-76.0)(0.45,80.9-76.0)(0.5,81.0-76.0)(0.55,81.2-76.0)(0.6,81.3-76.0)(0.65,81.6-76.0)(0.7,81.8-76.0)(0.75,82.0-76.0)(0.8,82.3-76.0)(0.85,82.2-76.0)(0.9,82.1-76.0)(0.95,81.6-76.0)(0.975,80.6-76.0)(0.9875,79.6-76.0)(0.99375,78.8-76.0)(1,77.0-76.0)};

\addplot[mark=none,line width=2pt,loosely dotted] coordinates {(0,0)(1,0)};

\nextgroupplot[title={\acs{knn}-based normalization on DCASE2020 evaluation set},xlabel=$K$,xtick={1,2,3,4,5,6,7,8,9,10,11,12,13,14,15},xticklabels={1,2,3,4,5,6,7,8,16,32,64,128,256,512,$\lvert\cX_\text{ref}\rvert$-1},xmin=1,xmax=15,xlabel style={yshift=4ex}]
% Direct-ACT
\addplot+[mark=square*,line width=2pt] coordinates {
(1,90.2-91.3)(2,90.2-91.3)(3,90.2-91.3)(4,90.2-91.3)(5,90.2-91.3)(6,90.2-91.3)(7,90.2-91.3)(8,90.2-91.3)(9,90.1-91.3)(10,90.0-91.3)(11,90.0-91.3)(12,90.0-91.3)(13,90.0-91.3)(14,90.2-91.3)(15,90.9-91.3)};

% OpenL3-raw
\addplot+[mark=triangle*,line width=2pt] coordinates {
(1,77.5-70.8)(2,77.8-70.8)(3,77.9-70.8)(4,78.0-70.8)(5,78.0-70.8)(6,78.1-70.8)(7,78.2-70.8)(8,78.3-70.8)(9,78.1-70.8)(10,77.2-70.8)(11,76.0-70.8)(12,74.9-70.8)(13,74.4-70.8)(14,74.3-70.8)(15,73.5-70.8)};

% BEATs-raw
\addplot+[mark=pentagon*,line width=2pt] coordinates {
(1,82.2-78.0)(2,83.2-78.0)(3,83.7-78.0)(4,83.9-78.0)(5,84.0-78.0)(6,84.1-78.0)(7,84.2-78.0)(8,84.2-78.0)(9,83.9-78.0)(10,82.9-78.0)(11,81.6-78.0)(12,80.2-78.0)(13,79.4-78.0)(14,79.4-78.0)(15,79.2-78.0)};

% EAT-raw
\addplot+[mark=*,line width=2pt] coordinates {
(1,79.9-76.0)(2,80.7-76.0)(3,81.1-76.0)(4,81.4-76.0)(5,81.7-76.0)(6,81.9-76.0)(7,82.1-76.0)(8,82.2-76.0)(9,82.2-76.0)(10,81.5-76.0)(11,80.6-76.0)(12,79.2-76.0)(13,78.4-76.0)(14,77.7-76.0)(15,77.0-76.0)};

\addplot[mark=none,line width=2pt,loosely dotted] coordinates {(1,0)(15,0)};

\nextgroupplot[title={\acs{gwrp}-based normalization on DCASE2022 development set},xlabel=$r$,
    xmin=0.0,
    xmax=1.0,]
% Direct-ACT
\addplot+[mark=square*,line width=2pt] coordinates {
(0,67.7-68.4)(0.05,67.7-68.4)(0.1,67.7-68.4)(0.15,67.8-68.4)(0.2,67.7-68.4)(0.25,67.7-68.4)(0.3,67.7-68.4)(0.35,67.6-68.4)(0.4,67.6-68.4)(0.45,67.5-68.4)(0.5,67.4-68.4)(0.55,67.3-68.4)(0.6,67.2-68.4)(0.65,67.0-68.4)(0.7,66.8-68.4)(0.75,66.5-68.4)(0.8,66.1-68.4)(0.85,65.7-68.4)(0.9,65.3-68.4)(0.95,65.1-68.4)(0.975,65.5-68.4)(0.9875,67.1-68.4)(0.99375,68.8-68.4)(1,68.4-68.4)};

% OpenL3-raw
\addplot+[mark=triangle*,line width=2pt] coordinates {
(0,59.3-59.4)(0.05,59.3-59.4)(0.1,59.3-59.4)(0.15,59.4-59.4)(0.2,59.3-59.4)(0.25,59.3-59.4)(0.3,59.3-59.4)(0.35,59.3-59.4)(0.4,59.3-59.4)(0.45,59.3-59.4)(0.5,59.2-59.4)(0.55,59.2-59.4)(0.6,59.1-59.4)(0.65,59.1-59.4)(0.7,59.2-59.4)(0.75,59.1-59.4)(0.8,59.1-59.4)(0.85,59.1-59.4)(0.9,59.1-59.4)(0.95,59.4-59.4)(0.975,59.7-59.4)(0.9875,59.9-59.4)(0.99375,59.8-59.4)(1,59.6-59.4)};

% BEATs-raw
\addplot+[mark=pentagon*,line width=2pt] coordinates {
(0,62.4-62.2)(0.05,62.5-62.2)(0.1,62.5-62.2)(0.15,62.5-62.2)(0.2,62.6-62.2)(0.25,62.6-62.2)(0.3,62.6-62.2)(0.35,62.7-62.2)(0.4,62.7-62.2)(0.45,62.7-62.2)(0.5,62.7-62.2)(0.55,62.7-62.2)(0.6,62.6-62.2)(0.65,62.5-62.2)(0.7,62.4-62.2)(0.75,62.3-62.2)(0.8,62.0-62.2)(0.85,61.6-62.2)(0.9,61.3-62.2)(0.95,61.0-62.2)(0.975,61.2-62.2)(0.9875,61.3-62.2)(0.99375,61.4-62.2)(1,62.1-62.2)};

% EAT-raw
\addplot+[mark=*,line width=2pt] coordinates {
(0,61.2-60.8)(0.05,61.3-60.8)(0.1,61.3-60.8)(0.15,61.4-60.8)(0.2,61.4-60.8)(0.25,61.5-60.8)(0.3,61.5-60.8)(0.35,61.5-60.8)(0.4,61.5-60.8)(0.45,61.6-60.8)(0.5,61.6-60.8)(0.55,61.7-60.8)(0.6,61.7-60.8)(0.65,61.6-60.8)(0.7,61.5-60.8)(0.75,61.4-60.8)(0.8,61.2-60.8)(0.85,61.2-60.8)(0.9,61.1-60.8)(0.95,60.9-60.8)(0.975,60.7-60.8)(0.9875,60.9-60.8)(0.99375,61.0-60.8)(1,60.8-60.8)};

\addplot[mark=none,line width=2pt,loosely dotted] coordinates {(0,0)(1,0)};

\nextgroupplot[title={\acs{knn}-based normalization on DCASE2022 development set},xlabel=$K$,xtick={1,2,3,4,5,6,7,8,9,10,11,12,13,14,15},xticklabels={1,2,3,4,5,6,7,8,16,32,64,128,256,512,$\lvert\cX_\text{ref}\rvert$-1},xmin=1,xmax=15,xlabel style={yshift=4ex}]
% Direct-ACT
\addplot+[mark=square*,line width=2pt] coordinates {
(1,67.7-68.4)(2,67.7-68.4)(3,67.4-68.4)(4,67.1-68.4)(5,66.8-68.4)(6,66.5-68.4)(7,66.2-68.4)(8,65.9-68.4)(9,64.9-68.4)(10,64.8-68.4)(11,65.1-68.4)(12,65.7-68.4)(13,67.3-68.4)(14,68.7-68.4)(15,68.4-68.4)};

% OpenL3-raw
\addplot+[mark=triangle*,line width=2pt] coordinates {
(1,59.3-59.4)(2,59.4-59.4)(3,59.3-59.4)(4,59.2-59.4)(5,59.2-59.4)(6,59.3-59.4)(7,59.3-59.4)(8,59.3-59.4)(9,59.1-59.4)(10,59.3-59.4)(11,59.6-59.4)(12,59.7-59.4)(13,59.6-59.4)(14,59.3-59.4)(15,59.6-59.4)};

% BEATs-raw
\addplot+[mark=pentagon*,line width=2pt] coordinates {
(1,62.4-62.2)(2,62.8-62.2)(3,62.7-62.2)(4,62.5-62.2)(5,62.4-62.2)(6,62.3-62.2)(7,62.1-62.2)(8,61.9-62.2)(9,61.0-62.2)(10,60.7-62.2)(11,60.8-62.2)(12,61.1-62.2)(13,61.1-62.2)(14,61.2-62.2)(15,62.1-62.2)};

% EAT-raw
\addplot+[mark=*,line width=2pt] coordinates {
(1,61.2-60.8)(2,61.5-60.8)(3,61.8-60.8)(4,61.7-60.8)(5,61.5-60.8)(6,61.4-60.8)(7,61.3-60.8)(8,61.3-60.8)(9,60.9-60.8)(10,60.7-60.8)(11,60.4-60.8)(12,60.6-60.8)(13,60.9-60.8)(14,60.7-60.8)(15,60.8-60.8)};

\addplot[mark=none,line width=2pt,loosely dotted] coordinates {(1,0)(15,0)};

\nextgroupplot[title={\acs{gwrp}-based normalization on DCASE2022 evaluation set},xlabel=$r$,
    xmin=0.0,
    xmax=1.0,]
% Direct-ACT
\addplot+[mark=square*,line width=2pt] coordinates {
(0,66.1-64.0)(0.05,66.2-64.0)(0.1,66.3-64.0)(0.15,66.3-64.0)(0.2,66.4-64.0)(0.25,66.4-64.0)(0.3,66.4-64.0)(0.35,66.3-64.0)(0.4,66.2-64.0)(0.45,66.1-64.0)(0.5,65.9-64.0)(0.55,65.7-64.0)(0.6,65.5-64.0)(0.65,65.2-64.0)(0.7,64.9-64.0)(0.75,64.6-64.0)(0.8,64.2-64.0)(0.85,63.8-64.0)(0.9,63.5-64.0)(0.95,63.3-64.0)(0.975,63.8-64.0)(0.9875,66.1-64.0)(0.99375,67.6-64.0)(1,64.4-64.0)};

% OpenL3-raw
\addplot+[mark=triangle*,line width=2pt] coordinates {
(0,61.5-57.7)(0.05,61.6-57.7)(0.1,61.6-57.7)(0.15,61.6-57.7)(0.2,61.6-57.7)(0.25,61.6-57.7)(0.3,61.6-57.7)(0.35,61.5-57.7)(0.4,61.4-57.7)(0.45,61.3-57.7)(0.5,61.2-57.7)(0.55,61.1-57.7)(0.6,61.0-57.7)(0.65,60.8-57.7)(0.7,60.7-57.7)(0.75,60.5-57.7)(0.8,60.3-57.7)(0.85,60.2-57.7)(0.9,60.0-57.7)(0.95,59.9-57.7)(0.975,59.9-57.7)(0.9875,59.7-57.7)(0.99375,59.5-57.7)(1,58.6-57.7)};

% BEATs-raw
\addplot+[mark=pentagon*,line width=2pt] coordinates {
(0,62.3-60.5)(0.05,62.4-60.5)(0.1,62.4-60.5)(0.15,62.4-60.5)(0.2,62.4-60.5)(0.25,62.5-60.5)(0.3,62.5-60.5)(0.35,62.6-60.5)(0.4,62.6-60.5)(0.45,62.5-60.5)(0.5,62.5-60.5)(0.55,62.5-60.5)(0.6,62.4-60.5)(0.65,62.4-60.5)(0.7,62.3-60.5)(0.75,62.3-60.5)(0.8,62.2-60.5)(0.85,62.2-60.5)(0.9,62.0-60.5)(0.95,62.0-60.5)(0.975,61.9-60.5)(0.9875,61.9-60.5)(0.99375,61.9-60.5)(1,61.3-60.5)};

% EAT-raw
\addplot+[mark=*,line width=2pt] coordinates {
(0,61.7-59.1)(0.05,61.7-59.1)(0.1,61.6-59.1)(0.15,61.7-59.1)(0.2,61.7-59.1)(0.25,61.7-59.1)(0.3,61.8-59.1)(0.35,61.8-59.1)(0.4,61.8-59.1)(0.45,61.7-59.1)(0.5,61.7-59.1)(0.55,61.8-59.1)(0.6,61.9-59.1)(0.65,61.9-59.1)(0.7,61.9-59.1)(0.75,61.8-59.1)(0.8,61.8-59.1)(0.85,61.8-59.1)(0.9,61.8-59.1)(0.95,61.8-59.1)(0.975,61.7-59.1)(0.9875,61.8-59.1)(0.99375,61.7-59.1)(1,60.5-59.1)};

\addplot[mark=none,line width=2pt,loosely dotted] coordinates {(0,0)(1,0)};

\nextgroupplot[title={\acs{knn}-based normalization on DCASE2022 evaluation set},xlabel=$K$,xtick={1,2,3,4,5,6,7,8,9,10,11,12,13,14,15},xticklabels={1,2,3,4,5,6,7,8,16,32,64,128,256,512,$\lvert\cX_\text{ref}\rvert$-1},xmin=1,xmax=15,xlabel style={yshift=4ex}]
% Direct-ACT
\addplot+[mark=square*,line width=2pt] coordinates {
(1,66.1-64.0)(2,66.2-64.0)(3,65.7-64.0)(4,65.1-64.0)(5,64.7-64.0)(6,64.3-64.0)(7,64.0-64.0)(8,63.8-64.0)(9,63.0-64.0)(10,62.9-64.0)(11,63.0-64.0)(12,63.9-64.0)(13,66.8-64.0)(14,66.8-64.0)(15,64.4-64.0)};

% OpenL3-raw
\addplot+[mark=triangle*,line width=2pt] coordinates {
(1,61.5-57.7)(2,61.4-57.7)(3,61.1-57.7)(4,60.8-57.7)(5,60.5-57.7)(6,60.4-57.7)(7,60.3-57.7)(8,60.1-57.7)(9,59.8-57.7)(10,59.7-57.7)(11,59.9-57.7)(12,59.7-57.7)(13,59.3-57.7)(14,58.6-57.7)(15,58.6-57.7)};

% BEATs-raw
\addplot+[mark=pentagon*,line width=2pt] coordinates {
(1,62.3-60.5)(2,62.6-60.5)(3,62.5-60.5)(4,62.3-60.5)(5,62.1-60.5)(6,62.0-60.5)(7,62.0-60.5)(8,62.0-60.5)(9,61.7-60.5)(10,61.8-60.5)(11,61.8-60.5)(12,61.7-60.5)(13,62.0-60.5)(14,61.5-60.5)(15,61.3-60.5)};

% EAT-raw
\addplot+[mark=*,line width=2pt] coordinates {
(1,61.7-59.1)(2,61.8-59.1)(3,61.9-59.1)(4,61.8-59.1)(5,61.6-59.1)(6,61.5-59.1)(7,61.4-59.1)(8,61.5-59.1)(9,61.6-59.1)(10,61.6-59.1)(11,61.6-59.1)(12,61.5-59.1)(13,61.7-59.1)(14,61.1-59.1)(15,60.5-59.1)};

\addplot[mark=none,line width=2pt,loosely dotted] coordinates {(1,0)(15,0)};

\nextgroupplot[title={\acs{gwrp}-based normalization on DCASE2023 development set},xlabel=$r$,
    xmin=0.0,
    xmax=1.0,]
% Direct-ACT
\addplot+[mark=square*,line width=2pt] coordinates {
(0,68.4-65.2)(0.05,68.4-65.2)(0.1,68.4-65.2)(0.15,68.5-65.2)(0.2,68.5-65.2)(0.25,68.5-65.2)(0.3,68.5-65.2)(0.35,68.5-65.2)(0.4,68.6-65.2)(0.45,68.6-65.2)(0.5,68.7-65.2)(0.55,68.7-65.2)(0.6,68.7-65.2)(0.65,68.8-65.2)(0.7,68.8-65.2)(0.75,68.8-65.2)(0.8,68.7-65.2)(0.85,68.7-65.2)(0.9,68.7-65.2)(0.95,69.2-65.2)(0.975,70.0-65.2)(0.9875,69.6-65.2)(0.99375,68.7-65.2)(1,64.1-65.2)};

% OpenL3-raw
\addplot+[mark=triangle*,line width=2pt] coordinates {
(0,60.5-59.1)(0.05,60.5-59.1)(0.1,60.4-59.1)(0.15,60.4-59.1)(0.2,60.3-59.1)(0.25,60.2-59.1)(0.3,60.2-59.1)(0.35,60.2-59.1)(0.4,60.1-59.1)(0.45,60.1-59.1)(0.5,60.1-59.1)(0.55,60.0-59.1)(0.6,59.9-59.1)(0.65,59.9-59.1)(0.7,59.8-59.1)(0.75,59.8-59.1)(0.8,59.8-59.1)(0.85,59.6-59.1)(0.9,59.7-59.1)(0.95,60.4-59.1)(0.975,61.4-59.1)(0.9875,61.7-59.1)(0.99375,61.5-59.1)(1,60.3-59.1)};

% BEATs-raw
\addplot+[mark=pentagon*,line width=2pt] coordinates {
(0,64.7-62.5)(0.05,64.7-62.5)(0.1,64.7-62.5)(0.15,64.7-62.5)(0.2,64.7-62.5)(0.25,64.7-62.5)(0.3,64.7-62.5)(0.35,64.7-62.5)(0.4,64.8-62.5)(0.45,64.7-62.5)(0.5,64.8-62.5)(0.55,64.8-62.5)(0.6,64.7-62.5)(0.65,64.5-62.5)(0.7,64.2-62.5)(0.75,64.0-62.5)(0.8,63.5-62.5)(0.85,63.2-62.5)(0.9,62.7-62.5)(0.95,62.1-62.5)(0.975,62.4-62.5)(0.9875,62.3-62.5)(0.99375,62.4-62.5)(1,62.7-62.5)};

% EAT-raw
\addplot+[mark=*,line width=2pt] coordinates {
(0,65.0-60.8)(0.05,65.1-60.8)(0.1,65.1-60.8)(0.15,65.2-60.8)(0.2,65.2-60.8)(0.25,65.2-60.8)(0.3,65.2-60.8)(0.35,65.2-60.8)(0.4,65.2-60.8)(0.45,65.2-60.8)(0.5,65.1-60.8)(0.55,65.0-60.8)(0.6,65.1-60.8)(0.65,64.9-60.8)(0.7,64.7-60.8)(0.75,64.4-60.8)(0.8,64.1-60.8)(0.85,63.7-60.8)(0.9,63.4-60.8)(0.95,62.7-60.8)(0.975,62.5-60.8)(0.9875,62.7-60.8)(0.99375,62.7-60.8)(1,61.7-60.8)};

\addplot[mark=none,line width=2pt,loosely dotted] coordinates {(0,0)(1,0)};

\nextgroupplot[title={\acs{knn}-based normalization on DCASE2023 development set},xlabel=$K$,xtick={1,2,3,4,5,6,7,8,9,10,11,12,13,14,15},xticklabels={1,2,3,4,5,6,7,8,16,32,64,128,256,512,$\lvert\cX_\text{ref}\rvert$-1},xmin=1,xmax=15,xlabel style={yshift=4ex}]
% Direct-ACT
\addplot+[mark=square*,line width=2pt] coordinates {
(1,68.4-65.2)(2,68.7-65.2)(3,68.9-65.2)(4,68.9-65.2)(5,68.8-65.2)(6,68.8-65.2)(7,68.7-65.2)(8,68.6-65.2)(9,68.5-65.2)(10,68.7-65.2)(11,69.3-65.2)(12,69.6-65.2)(13,68.4-65.2)(14,67.2-65.2)(15,64.1-65.2)};

% OpenL3-raw
\addplot+[mark=triangle*,line width=2pt] coordinates {
(1,60.5-59.1)(2,60.1-59.1)(3,60.1-59.1)(4,60.0-59.1)(5,59.8-59.1)(6,59.7-59.1)(7,59.6-59.1)(8,59.5-59.1)(9,59.3-59.1)(10,60.0-59.1)(11,61.2-59.1)(12,61.6-59.1)(13,61.2-59.1)(14,61.0-59.1)(15,60.3-59.1)};

% BEATs-raw
\addplot+[mark=pentagon*,line width=2pt] coordinates {
(1,64.7-62.5)(2,65.2-62.5)(3,65.1-62.5)(4,64.9-62.5)(5,64.3-62.5)(6,63.9-62.5)(7,63.6-62.5)(8,63.3-62.5)(9,62.6-62.5)(10,62.0-62.5)(11,62.0-62.5)(12,62.1-62.5)(13,62.2-62.5)(14,62.3-62.5)(15,62.7-62.5)};

% EAT-raw
\addplot+[mark=*,line width=2pt] coordinates {
(1,65.0-60.8)(2,65.2-60.8)(3,65.2-60.8)(4,64.9-60.8)(5,64.4-60.8)(6,64.1-60.8)(7,63.9-60.8)(8,63.7-60.8)(9,62.9-60.8)(10,62.4-60.8)(11,62.2-60.8)(12,62.6-60.8)(13,62.6-60.8)(14,62.2-60.8)(15,61.7-60.8)};

\addplot[mark=none,line width=2pt,loosely dotted] coordinates {(1,0)(15,0)};

\nextgroupplot[title={\acs{gwrp}-based normalization on DCASE2023 evaluation set},xlabel=$r$,
    xmin=0.0,
    xmax=1.0,]
% Direct-ACT
\addplot+[mark=square*,line width=2pt] coordinates {
(0,68.0-56.2)(0.05,68.0-56.2)(0.1,68.1-56.2)(0.15,68.1-56.2)(0.2,68.1-56.2)(0.25,68.1-56.2)(0.3,68.0-56.2)(0.35,67.9-56.2)(0.4,67.8-56.2)(0.45,67.7-56.2)(0.5,67.6-56.2)(0.55,67.4-56.2)(0.6,67.2-56.2)(0.65,66.9-56.2)(0.7,66.6-56.2)(0.75,66.1-56.2)(0.8,65.6-56.2)(0.85,65.1-56.2)(0.9,64.8-56.2)(0.95,64.4-56.2)(0.975,64.8-56.2)(0.9875,66.3-56.2)(0.99375,66.9-56.2)(1,57.7-56.2)};

% OpenL3-raw
\addplot+[mark=triangle*,line width=2pt] coordinates {
(0,63.8-59.0)(0.05,63.8-59.0)(0.1,63.8-59.0)(0.15,63.7-59.0)(0.2,63.7-59.0)(0.25,63.6-59.0)(0.3,63.5-59.0)(0.35,63.5-59.0)(0.4,63.5-59.0)(0.45,63.5-59.0)(0.5,63.3-59.0)(0.55,63.1-59.0)(0.6,62.7-59.0)(0.65,62.3-59.0)(0.7,61.9-59.0)(0.75,61.5-59.0)(0.8,60.9-59.0)(0.85,60.3-59.0)(0.9,59.7-59.0)(0.95,59.0-59.0)(0.975,58.8-59.0)(0.9875,58.9-59.0)(0.99375,59.8-59.0)(1,60.7-59.0)};

% BEATs-raw
\addplot+[mark=pentagon*,line width=2pt] coordinates {
(0,67.6-59.0)(0.05,67.8-59.0)(0.1,67.9-59.0)(0.15,68.0-59.0)(0.2,68.2-59.0)(0.25,68.2-59.0)(0.3,68.2-59.0)(0.35,68.2-59.0)(0.4,68.2-59.0)(0.45,68.1-59.0)(0.5,68.0-59.0)(0.55,67.6-59.0)(0.6,67.3-59.0)(0.65,66.9-59.0)(0.7,66.3-59.0)(0.75,65.6-59.0)(0.8,64.5-59.0)(0.85,63.6-59.0)(0.9,63.2-59.0)(0.95,62.9-59.0)(0.975,62.5-59.0)(0.9875,62.2-59.0)(0.99375,61.5-59.0)(1,60.6-59.0)};

% EAT-raw
\addplot+[mark=*,line width=2pt] coordinates {
(0,66.1-59.5)(0.05,66.2-59.5)(0.1,66.2-59.5)(0.15,66.3-59.5)(0.2,66.4-59.5)(0.25,66.4-59.5)(0.3,66.4-59.5)(0.35,66.4-59.5)(0.4,66.2-59.5)(0.45,65.9-59.5)(0.5,65.7-59.5)(0.55,65.4-59.5)(0.6,64.9-59.5)(0.65,64.0-59.5)(0.7,63.0-59.5)(0.75,62.3-59.5)(0.8,61.6-59.5)(0.85,60.9-59.5)(0.9,60.4-59.5)(0.95,60.0-59.5)(0.975,59.8-59.5)(0.9875,59.5-59.5)(0.99375,59.1-59.5)(1,59.3-59.5)};

\addplot[mark=none,line width=2pt,loosely dotted] coordinates {(0,0)(1,0)};

\nextgroupplot[title={\acs{knn}-based normalization on DCASE2023 evaluation set},xlabel=$K$,xtick={1,2,3,4,5,6,7,8,9,10,11,12,13,14,15},xticklabels={1,2,3,4,5,6,7,8,16,32,64,128,256,512,$\lvert\cX_\text{ref}\rvert$-1},xmin=1,xmax=15,xlabel style={yshift=4ex}]
% Direct-ACT
\addplot+[mark=square*,line width=2pt] coordinates {
(1,68.0-56.2)(2,67.8-56.2)(3,67.4-56.2)(4,67.1-56.2)(5,66.6-56.2)(6,66.2-56.2)(7,65.9-56.2)(8,65.5-56.2)(9,64.4-56.2)(10,64.2-56.2)(11,64.2-56.2)(12,64.7-56.2)(13,65.8-56.2)(14,65.4-56.2)(15,57.7-56.2)};

% OpenL3-raw
\addplot+[mark=triangle*,line width=2pt] coordinates {
(1,63.8-59.0)(2,63.4-59.0)(3,63.0-59.0)(4,62.5-59.0)(5,61.9-59.0)(6,61.5-59.0)(7,61.2-59.0)(8,60.8-59.0)(9,59.6-59.0)(10,58.9-59.0)(11,58.5-59.0)(12,58.7-59.0)(13,59.3-59.0)(14,60.2-59.0)(15,60.7-59.0)};

% BEATs-raw
\addplot+[mark=pentagon*,line width=2pt] coordinates {
(1,67.6-59.0)(2,68.2-59.0)(3,67.9-59.0)(4,67.4-59.0)(5,67.0-59.0)(6,66.7-59.0)(7,66.3-59.0)(8,65.9-59.0)(9,63.0-59.0)(10,62.8-59.0)(11,62.3-59.0)(12,62.0-59.0)(13,61.5-59.0)(14,60.8-59.0)(15,60.6-59.0)};

% EAT-raw
\addplot+[mark=*,line width=2pt] coordinates {
(1,66.1-59.5)(2,66.0-59.5)(3,65.5-59.5)(4,64.4-59.5)(5,63.6-59.5)(6,63.1-59.5)(7,62.4-59.5)(8,62.0-59.5)(9,60.2-59.5)(10,59.9-59.5)(11,59.7-59.5)(12,59.6-59.5)(13,59.1-59.5)(14,58.9-59.5)(15,59.3-59.5)};

\addplot[mark=none,line width=2pt,loosely dotted] coordinates {(1,0)(15,0)};

\nextgroupplot[title={\acs{gwrp}-based normalization on DCASE2024 development set},xlabel=$r$,
    xmin=0.0,
    xmax=1.0,]
% Direct-ACT
\addplot+[mark=square*,line width=2pt] coordinates {
(0,62.0-58.1)(0.05,62.0-58.1)(0.1,62.0-58.1)(0.15,62.0-58.1)(0.2,62.0-58.1)(0.25,62.0-58.1)(0.3,62.0-58.1)(0.35,62.0-58.1)(0.4,62.0-58.1)(0.45,61.9-58.1)(0.5,61.9-58.1)(0.55,61.9-58.1)(0.6,61.8-58.1)(0.65,61.8-58.1)(0.7,61.7-58.1)(0.75,61.6-58.1)(0.8,61.5-58.1)(0.85,61.5-58.1)(0.9,61.3-58.1)(0.95,61.3-58.1)(0.975,61.2-58.1)(0.9875,60.5-58.1)(0.99375,60.0-58.1)(1,58.6-58.1)};

% OpenL3-raw
\addplot+[mark=triangle*,line width=2pt] coordinates {
(0,56.6-54.7)(0.05,56.6-54.7)(0.1,56.6-54.7)(0.15,56.6-54.7)(0.2,56.6-54.7)(0.25,56.6-54.7)(0.3,56.6-54.7)(0.35,56.6-54.7)(0.4,56.6-54.7)(0.45,56.5-54.7)(0.5,56.4-54.7)(0.55,56.3-54.7)(0.6,56.2-54.7)(0.65,56.1-54.7)(0.7,56.1-54.7)(0.75,56.0-54.7)(0.8,56.0-54.7)(0.85,56.1-54.7)(0.9,56.2-54.7)(0.95,56.2-54.7)(0.975,56.4-54.7)(0.9875,56.7-54.7)(0.99375,56.8-54.7)(1,56.2-54.7)};

% BEATs-raw
\addplot+[mark=pentagon*,line width=2pt] coordinates {
(0,58.1-56.5)(0.05,58.2-56.5)(0.1,58.3-56.5)(0.15,58.5-56.5)(0.2,58.5-56.5)(0.25,58.6-56.5)(0.3,58.7-56.5)(0.35,58.7-56.5)(0.4,58.7-56.5)(0.45,58.8-56.5)(0.5,58.9-56.5)(0.55,58.9-56.5)(0.6,58.9-56.5)(0.65,58.9-56.5)(0.7,58.8-56.5)(0.75,58.7-56.5)(0.8,58.5-56.5)(0.85,58.4-56.5)(0.9,58.2-56.5)(0.95,57.9-56.5)(0.975,58.1-56.5)(0.9875,58.3-56.5)(0.99375,58.7-56.5)(1,58.6-56.5)};

% EAT-raw
\addplot+[mark=*,line width=2pt] coordinates {
(0,56.8-56.8)(0.05,56.8-56.8)(0.1,56.8-56.8)(0.15,56.8-56.8)(0.2,56.9-56.8)(0.25,56.9-56.8)(0.3,57.0-56.8)(0.35,56.9-56.8)(0.4,56.9-56.8)(0.45,57.0-56.8)(0.5,57.1-56.8)(0.55,57.0-56.8)(0.6,57.0-56.8)(0.65,57.1-56.8)(0.7,57.0-56.8)(0.75,57.0-56.8)(0.8,56.9-56.8)(0.85,57.0-56.8)(0.9,56.9-56.8)(0.95,56.8-56.8)(0.975,57.0-56.8)(0.9875,57.5-56.8)(0.99375,57.7-56.8)(1,57.1-56.8)};

\addplot[mark=none,line width=2pt,loosely dotted] coordinates {(0,0)(1,0)};

\nextgroupplot[title={\acs{knn}-based normalization on DCASE2024 development set},xlabel=$K$,xtick={1,2,3,4,5,6,7,8,9,10,11,12,13,14,15},xticklabels={1,2,3,4,5,6,7,8,16,32,64,128,256,512,$\lvert\cX_\text{ref}\rvert$-1},xmin=1,xmax=15,xlabel style={yshift=4ex}]
% Direct-ACT
\addplot+[mark=square*,line width=2pt] coordinates {
(1,62.0-58.1)(2,62.0-58.1)(3,61.9-58.1)(4,61.8-58.1)(5,61.7-58.1)(6,61.6-58.1)(7,61.5-58.1)(8,61.5-58.1)(9,61.2-58.1)(10,61.1-58.1)(11,61.2-58.1)(12,60.6-58.1)(13,59.8-58.1)(14,59.0-58.1)(15,58.6-58.1)};

% OpenL3-raw
\addplot+[mark=triangle*,line width=2pt] coordinates {
(1,56.6-54.7)(2,56.5-54.7)(3,56.4-54.7)(4,56.2-54.7)(5,56.0-54.7)(6,55.9-54.7)(7,55.9-54.7)(8,55.9-54.7)(9,55.9-54.7)(10,55.9-54.7)(11,56.2-54.7)(12,56.4-54.7)(13,56.7-54.7)(14,56.7-54.7)(15,56.2-54.7)};

% BEATs-raw
\addplot+[mark=pentagon*,line width=2pt] coordinates {
(1,58.1-56.5)(2,59.0-56.5)(3,59.2-56.5)(4,59.0-56.5)(5,58.9-56.5)(6,58.7-56.5)(7,58.6-56.5)(8,58.5-56.5)(9,58.1-56.5)(10,57.9-56.5)(11,57.8-56.5)(12,58.2-56.5)(13,58.5-56.5)(14,58.8-56.5)(15,58.5-56.5)};

% EAT-raw
\addplot+[mark=*,line width=2pt] coordinates {
(1,56.8-56.8)(2,56.9-56.8)(3,57.1-56.8)(4,57.2-56.8)(5,57.1-56.8)(6,56.9-56.8)(7,57.0-56.8)(8,57.0-56.8)(9,56.8-56.8)(10,56.7-56.8)(11,57.0-56.8)(12,57.3-56.8)(13,57.6-56.8)(14,57.5-56.8)(15,57.1-56.8)};

\addplot[mark=none,line width=2pt,loosely dotted] coordinates {(1,0)(15,0)};

\nextgroupplot[title={\acs{gwrp}-based normalization on DCASE2024 evaluation set},xlabel=$r$,
    xmin=0.0,
    xmax=1.0,]
% Direct-ACT
\addplot+[mark=square*,line width=2pt] coordinates {
(0,54.7-53.0)(0.05,54.7-53.0)(0.1,54.6-53.0)(0.15,54.6-53.0)(0.2,54.5-53.0)(0.25,54.5-53.0)(0.3,54.4-53.0)(0.35,54.3-53.0)(0.4,54.2-53.0)(0.45,54.1-53.0)(0.5,54.0-53.0)(0.55,53.7-53.0)(0.6,53.4-53.0)(0.65,52.9-53.0)(0.7,52.3-53.0)(0.75,51.5-53.0)(0.8,50.8-53.0)(0.85,50.4-53.0)(0.9,50.4-53.0)(0.95,51.3-53.0)(0.975,52.7-53.0)(0.9875,53.5-53.0)(0.99375,53.7-53.0)(1,52.7-53.0)};

% OpenL3-raw
\addplot+[mark=triangle*,line width=2pt] coordinates {
(0,55.7-55.9)(0.05,55.7-55.9)(0.1,55.7-55.9)(0.15,55.7-55.9)(0.2,55.7-55.9)(0.25,55.6-55.9)(0.3,55.6-55.9)(0.35,55.4-55.9)(0.4,55.3-55.9)(0.45,55.1-55.9)(0.5,55.0-55.9)(0.55,54.9-55.9)(0.6,54.6-55.9)(0.65,54.4-55.9)(0.7,54.0-55.9)(0.75,53.5-55.9)(0.8,52.9-55.9)(0.85,52.2-55.9)(0.9,51.7-55.9)(0.95,51.3-55.9)(0.975,51.6-55.9)(0.9875,52.6-55.9)(0.99375,53.5-55.9)(1,55.3-55.9)};

% BEATs-raw
\addplot+[mark=pentagon*,line width=2pt] coordinates {
(0,62.4-58.2)(0.05,62.2-58.2)(0.1,62.1-58.2)(0.15,61.9-58.2)(0.2,61.7-58.2)(0.25,61.4-58.2)(0.3,61.0-58.2)(0.35,60.7-58.2)(0.4,60.3-58.2)(0.45,59.9-58.2)(0.5,59.6-58.2)(0.55,59.2-58.2)(0.6,58.9-58.2)(0.65,58.3-58.2)(0.7,57.6-58.2)(0.75,56.7-58.2)(0.8,55.9-58.2)(0.85,55.2-58.2)(0.9,54.7-58.2)(0.95,54.3-58.2)(0.975,54.3-58.2)(0.9875,54.8-58.2)(0.99375,55.2-58.2)(1,55.4-58.2)};

% EAT-raw
\addplot+[mark=*,line width=2pt] coordinates {
(0,58.9-57.6)(0.05,58.9-57.6)(0.1,58.9-57.6)(0.15,58.9-57.6)(0.2,58.9-57.6)(0.25,58.8-57.6)(0.3,58.7-57.6)(0.35,58.5-57.6)(0.4,58.3-57.6)(0.45,58.1-57.6)(0.5,57.8-57.6)(0.55,57.7-57.6)(0.6,57.4-57.6)(0.65,57.0-57.6)(0.7,56.4-57.6)(0.75,55.9-57.6)(0.8,55.2-57.6)(0.85,54.5-57.6)(0.9,54.1-57.6)(0.95,54.0-57.6)(0.975,54.5-57.6)(0.9875,54.9-57.6)(0.99375,55.3-57.6)(1,56.1-57.6)};

\addplot[mark=none,line width=2pt,loosely dotted] coordinates {(0,0)(1,0)};

\nextgroupplot[title={\acs{knn}-based normalization on DCASE2024 evaluation set},xlabel=$K$,xtick={1,2,3,4,5,6,7,8,9,10,11,12,13,14,15},xticklabels={1,2,3,4,5,6,7,8,16,32,64,128,256,512,$\lvert\cX_\text{ref}\rvert$-1},xmin=1,xmax=15,xlabel style={yshift=4ex}]
% Direct-ACT
\addplot+[mark=square*,line width=2pt] coordinates {
(1,54.7-53.0)(2,54.3-53.0)(3,53.9-53.0)(4,53.4-53.0)(5,53.0-53.0)(6,52.6-53.0)(7,52.2-53.0)(8,51.8-53.0)(9,50.2-53.0)(10,50.5-53.0)(11,51.9-53.0)(12,53.5-53.0)(13,53.5-53.0)(14,53.3-53.0)(15,52.7-53.0)};

% OpenL3-raw
\addplot+[mark=triangle*,line width=2pt] coordinates {
(1,55.7-55.9)(2,55.1-55.9)(3,54.8-55.9)(4,54.5-55.9)(5,54.3-55.9)(6,53.9-55.9)(7,53.6-55.9)(8,53.0-55.9)(9,51.2-55.9)(10,51.0-55.9)(11,51.4-55.9)(12,52.4-55.9)(13,53.4-55.9)(14,54.3-55.9)(15,55.3-55.9)};

% BEATs-raw
\addplot+[mark=pentagon*,line width=2pt] coordinates {
(1,62.4-58.2)(2,61.0-58.2)(3,59.2-58.2)(4,58.6-58.2)(5,58.1-58.2)(6,57.6-58.2)(7,57.3-58.2)(8,57.0-58.2)(9,54.3-58.2)(10,54.0-58.2)(11,54.2-58.2)(12,54.7-58.2)(13,55.1-58.2)(14,55.5-58.2)(15,55.4-58.2)};

% EAT-raw
\addplot+[mark=*,line width=2pt] coordinates {
(1,58.9-57.6)(2,59.0-57.6)(3,58.0-57.6)(4,57.2-57.6)(5,57.2-57.6)(6,56.9-57.6)(7,56.6-57.6)(8,56.2-57.6)(9,53.7-57.6)(10,53.8-57.6)(11,54.3-57.6)(12,54.7-57.6)(13,55.1-57.6)(14,55.6-57.6)(15,56.1-57.6)};

\addplot[mark=none,line width=2pt,loosely dotted] coordinates {(1,0)(15,0)};

\nextgroupplot[title={\acs{gwrp}-based normalization on DCASE2025 development set},xlabel=$r$,
    xmin=0.0,
    xmax=1.0,]
% Direct-ACT
\addplot+[mark=square*,line width=2pt] coordinates {
(0,56.1-55.7)(0.05,56.1-55.7)(0.1,56.1-55.7)(0.15,56.1-55.7)(0.2,56.1-55.7)(0.25,56.1-55.7)(0.3,56.1-55.7)(0.35,56.1-55.7)(0.4,56.1-55.7)(0.45,56.0-55.7)(0.5,56.0-55.7)(0.55,56.0-55.7)(0.6,56.0-55.7)(0.65,55.9-55.7)(0.7,55.9-55.7)(0.75,55.9-55.7)(0.8,55.8-55.7)(0.85,55.8-55.7)(0.9,56.0-55.7)(0.95,57.2-55.7)(0.975,58.2-55.7)(0.9875,59.0-55.7)(0.99375,59.4-55.7)(1,57.2-55.7)};

% OpenL3-raw
\addplot+[mark=triangle*,line width=2pt] coordinates {
(0,56.0-54.6)(0.05,56.0-54.6)(0.1,56.0-54.6)(0.15,56.0-54.6)(0.2,56.1-54.6)(0.25,56.0-54.6)(0.3,56.0-54.6)(0.35,56.0-54.6)(0.4,56.0-54.6)(0.45,55.9-54.6)(0.5,55.9-54.6)(0.55,55.8-54.6)(0.6,55.7-54.6)(0.65,55.6-54.6)(0.7,55.5-54.6)(0.75,55.4-54.6)(0.8,55.3-54.6)(0.85,55.1-54.6)(0.9,55.4-54.6)(0.95,55.8-54.6)(0.975,55.9-54.6)(0.9875,55.9-54.6)(0.99375,55.9-54.6)(1,55.4-54.6)};

% BEATs-raw
\addplot+[mark=pentagon*,line width=2pt] coordinates {
(0,57.7-61.4)(0.05,57.7-61.4)(0.1,57.6-61.4)(0.15,57.5-61.4)(0.2,57.5-61.4)(0.25,57.4-61.4)(0.3,57.4-61.4)(0.35,57.4-61.4)(0.4,57.3-61.4)(0.45,57.3-61.4)(0.5,57.3-61.4)(0.55,57.1-61.4)(0.6,57.2-61.4)(0.65,57.1-61.4)(0.7,57.1-61.4)(0.75,57.1-61.4)(0.8,57.1-61.4)(0.85,57.1-61.4)(0.9,57.5-61.4)(0.95,59.0-61.4)(0.975,60.4-61.4)(0.9875,60.4-61.4)(0.99375,60.6-61.4)(1,60.9-61.4)};

% EAT-raw
\addplot+[mark=*,line width=2pt] coordinates {
(0,59.8-60.3)(0.05,59.8-60.3)(0.1,59.9-60.3)(0.15,59.8-60.3)(0.2,59.8-60.3)(0.25,59.8-60.3)(0.3,59.8-60.3)(0.35,59.9-60.3)(0.4,59.9-60.3)(0.45,59.9-60.3)(0.5,60.0-60.3)(0.55,60.0-60.3)(0.6,59.9-60.3)(0.65,60.0-60.3)(0.7,59.9-60.3)(0.75,59.8-60.3)(0.8,59.7-60.3)(0.85,59.7-60.3)(0.9,59.5-60.3)(0.95,60.0-60.3)(0.975,60.5-60.3)(0.9875,61.0-60.3)(0.99375,61.6-60.3)(1,61.2-60.3)};

\addplot[mark=none,line width=2pt,loosely dotted] coordinates {(0,0)(1,0)};

\nextgroupplot[title={\acs{knn}-based normalization on DCASE2025 development set},xlabel=$K$,xtick={1,2,3,4,5,6,7,8,9,10,11,12,13,14,15},xticklabels={1,2,3,4,5,6,7,8,16,32,64,128,256,512,$\lvert\cX_\text{ref}\rvert$-1},xmin=1,xmax=15,xlabel style={yshift=4ex}]
% Direct-ACT
\addplot+[mark=square*,line width=2pt] coordinates {
(1,56.1-55.7)(2,56.1-55.7)(3,56.1-55.7)(4,56.1-55.7)(5,55.9-55.7)(6,55.9-55.7)(7,55.9-55.7)(8,55.8-55.7)(9,55.7-55.7)(10,56.3-55.7)(11,58.0-55.7)(12,58.6-55.7)(13,58.8-55.7)(14,58.5-55.7)(15,57.2-55.7)};

% OpenL3-raw
\addplot+[mark=triangle*,line width=2pt] coordinates {
(1,56.0-54.6)(2,55.9-54.6)(3,55.9-54.6)(4,55.6-54.6)(5,55.5-54.6)(6,55.3-54.6)(7,55.2-54.6)(8,55.1-54.6)(9,55.0-54.6)(10,55.5-54.6)(11,55.7-54.6)(12,56.1-54.6)(13,55.7-54.6)(14,55.3-54.6)(15,55.4-54.6)};

% BEATs-raw
\addplot+[mark=pentagon*,line width=2pt] coordinates {
(1,57.7-61.4)(2,57.3-61.4)(3,57.3-61.4)(4,57.0-61.4)(5,56.9-61.4)(6,56.9-61.4)(7,56.8-61.4)(8,56.7-61.4)(9,57.0-61.4)(10,58.6-61.4)(11,60.4-61.4)(12,60.4-61.4)(13,60.3-61.4)(14,60.4-61.4)(15,60.9-61.4)};

% EAT-raw
\addplot+[mark=*,line width=2pt] coordinates {
(1,59.8-60.3)(2,60.0-60.3)(3,60.0-60.3)(4,59.9-60.3)(5,59.8-60.3)(6,59.5-60.3)(7,59.3-60.3)(8,59.2-60.3)(9,59.1-60.3)(10,59.5-60.3)(11,60.2-60.3)(12,61.0-60.3)(13,61.3-60.3)(14,61.7-60.3)(15,61.2-60.3)};

\addplot[mark=none,line width=2pt,loosely dotted] coordinates {(1,0)(15,0)};

\nextgroupplot[title={\acs{gwrp}-based normalization on DCASE2025 evaluation set},xlabel=$r$,
    xmin=0.0,
    xmax=1.0,]
% Direct-ACT
\addplot+[mark=square*,line width=2pt] coordinates {
(0,55.6-49.0)(0.05,55.6-49.0)(0.1,55.7-49.0)(0.15,55.7-49.0)(0.2,55.8-49.0)(0.25,55.8-49.0)(0.3,55.9-49.0)(0.35,55.9-49.0)(0.4,55.9-49.0)(0.45,55.9-49.0)(0.5,55.9-49.0)(0.55,55.9-49.0)(0.6,55.9-49.0)(0.65,55.9-49.0)(0.7,55.9-49.0)(0.75,55.7-49.0)(0.8,55.5-49.0)(0.85,55.2-49.0)(0.9,54.8-49.0)(0.95,54.4-49.0)(0.975,54.4-49.0)(0.9875,54.9-49.0)(0.99375,55.0-49.0)(1,52.5-49.0)};

% OpenL3-raw
\addplot+[mark=triangle*,line width=2pt] coordinates {
(0,57.3-54.0)(0.05,57.4-54.0)(0.1,57.4-54.0)(0.15,57.5-54.0)(0.2,57.6-54.0)(0.25,57.8-54.0)(0.3,57.8-54.0)(0.35,57.9-54.0)(0.4,58.0-54.0)(0.45,58.0-54.0)(0.5,58.0-54.0)(0.55,58.1-54.0)(0.6,58.1-54.0)(0.65,58.1-54.0)(0.7,58.0-54.0)(0.75,57.9-54.0)(0.8,57.7-54.0)(0.85,57.5-54.0)(0.9,57.3-54.0)(0.95,56.7-54.0)(0.975,56.2-54.0)(0.9875,55.9-54.0)(0.99375,55.6-54.0)(1,55.7-54.0)};

% BEATs-raw
\addplot+[mark=pentagon*,line width=2pt] coordinates {
(0,62.0-55.2)(0.05,62.0-55.2)(0.1,61.9-55.2)(0.15,62.0-55.2)(0.2,62.0-55.2)(0.25,62.0-55.2)(0.3,62.0-55.2)(0.35,62.0-55.2)(0.4,61.9-55.2)(0.45,62.0-55.2)(0.5,61.9-55.2)(0.55,61.8-55.2)(0.6,61.7-55.2)(0.65,61.7-55.2)(0.7,61.7-55.2)(0.75,61.6-55.2)(0.8,61.4-55.2)(0.85,61.2-55.2)(0.9,60.8-55.2)(0.95,60.0-55.2)(0.975,59.2-55.2)(0.9875,58.7-55.2)(0.99375,58.1-55.2)(1,56.9-55.2)};

% EAT-raw
\addplot+[mark=*,line width=2pt] coordinates {
(0,58.9-53.5)(0.05,59.0-53.5)(0.1,59.1-53.5)(0.15,59.1-53.5)(0.2,59.1-53.5)(0.25,59.1-53.5)(0.3,59.2-53.5)(0.35,59.2-53.5)(0.4,59.2-53.5)(0.45,59.2-53.5)(0.5,59.1-53.5)(0.55,59.2-53.5)(0.6,59.1-53.5)(0.65,59.0-53.5)(0.7,59.0-53.5)(0.75,59.0-53.5)(0.8,58.9-53.5)(0.85,58.8-53.5)(0.9,58.4-53.5)(0.95,57.9-53.5)(0.975,57.2-53.5)(0.9875,56.4-53.5)(0.99375,55.7-53.5)(1,55.0-53.5)};

\addplot[mark=none,line width=2pt,loosely dotted] coordinates {(0,0)(1,0)};

\nextgroupplot[title={\acs{knn}-based normalization on DCASE2025 evaluation set},xlabel=$K$,xtick={1,2,3,4,5,6,7,8,9,10,11,12,13,14,15},xticklabels={1,2,3,4,5,6,7,8,16,32,64,128,256,512,$\lvert\cX_\text{ref}\rvert$-1},xmin=1,xmax=15,xlabel style={yshift=4ex}]
% Direct-ACT
\addplot+[mark=square*,line width=2pt] coordinates {
(1,55.6-49.0)(2,55.9-49.0)(3,56.0-49.0)(4,56.1-49.0)(5,56.0-49.0)(6,55.9-49.0)(7,55.8-49.0)(8,55.6-49.0)(9,54.8-49.0)(10,54.3-49.0)(11,54.1-49.0)(12,54.2-49.0)(13,54.4-49.0)(14,54.7-49.0)(15,52.5-49.0)};

% OpenL3-raw
\addplot+[mark=triangle*,line width=2pt] coordinates {
(1,57.3-54.0)(2,57.9-54.0)(3,58.1-54.0)(4,58.1-54.0)(5,58.0-54.0)(6,57.9-54.0)(7,57.7-54.0)(8,57.6-54.0)(9,57.1-54.0)(10,56.6-54.0)(11,56.1-54.0)(12,55.8-54.0)(13,55.4-54.0)(14,55.1-54.0)(15,55.7-54.0)};

% BEATs-raw
\addplot+[mark=pentagon*,line width=2pt] coordinates {
(1,62.0-55.2)(2,62.0-55.2)(3,61.7-55.2)(4,61.6-55.2)(5,61.5-55.2)(6,61.5-55.2)(7,61.4-55.2)(8,61.3-55.2)(9,60.5-55.2)(10,59.7-55.2)(11,59.2-55.2)(12,58.6-55.2)(13,58.0-55.2)(14,57.0-55.2)(15,56.9-55.2)};

% EAT-raw
\addplot+[mark=*,line width=2pt] coordinates {
(1,58.9-53.5)(2,59.3-53.5)(3,59.1-53.5)(4,59.0-53.5)(5,59.0-53.5)(6,59.0-53.5)(7,58.8-53.5)(8,58.7-53.5)(9,58.3-53.5)(10,57.9-53.5)(11,57.2-53.5)(12,56.3-53.5)(13,55.7-53.5)(14,55.0-53.5)(15,55.0-53.5)};

\addplot[mark=none,line width=2pt,loosely dotted] coordinates {(1,0)(15,0)};

\end{groupplot}
\end{tikzpicture}

%% file: figs/sota.tex
\begin{table*}
\centering
\caption{Harmonic means of all AUCs and pAUCs obtained with different \acs{asd} systems on a representative group of \acs{asd} datasets used as benchmarks in several recent peer-reviewed works. For all of our proposed \acs{asd} systems, a ratio-based anomaly score normalization with $K=1$ was used. Whenever applicable, means of all independent trials are shown. Highest numbers in each column are in bold.}
\begin{adjustbox}{max width=\textwidth}
\begin{NiceTabular}{lcc*{8}{c}} %llllllll}
\toprule
&&\multicolumn{3}{c}{DCASE2020 dataset \cite{koizumi2020description}}&\multicolumn{3}{c}{DCASE2023 dataset \cite{dohi2023description}}&\multicolumn{3}{c}{DCASE2024 dataset \cite{nishida2024description}}\\
&&\multicolumn{3}{c}{(no domain shifts)}&\multicolumn{3}{c}{(first-shot \ac{dg})}&\multicolumn{3}{c}{(first-shot \ac{dg} with less meta data)}\\
\cmidrule(lr){3-5}\cmidrule(lr){6-8}\cmidrule(lr){9-11}
\acs{asd} system&trials&
dev. set&eval. set&arithm. mean&
dev. set&eval. set&harm. mean&
dev. set&eval. set&harm. mean\\
\midrule
Direct-\ac{act}&10&$90.7\%$&$90.2\%$&$90.5\%$&$68.4\%$&$68.0\%$&$68.2\%$&$62.0\%$&$54.7\%$&$58.1\%$\\
OpenL3-raw&1&$75.2\%$&$77.5\%$&$76.4\%$&$60.5\%$&$63.8\%$&$62.1\%$&$56.6\%$&$55.7\%$&$56.1\%$\\
BEATs-raw&1&$81.5\%$&$82.2\%$&$81.9\%$&$64.8\%$&$67.6\%$&$66.2\%$&$58.1\%$&$62.4\%$&$60.2\%$\\
EAT-raw&1&$79.3\%$&$79.9\%$&$79.6\%$&$65.0\%$&$66.1\%$&$65.5\%$&$56.8\%$&$58.9\%$&$57.8\%$\\
Direct-\ac{act} (ensemble)&1&$\pmb{94.2\%}$&$93.3\%$&$\pmb{93.8\%}$&$\pmb{71.3\%}$&$72.4\%$&$\pmb{71.8\%}$&$\pmb{65.2\%}$&$56.5\%$&$60.5\%$\\
\midrule
Koizumi et al. \cite{koizumi2020description}&1&$66.6\%$&$70.0\%$&$68.3\%$&$-$&$-$&$-$&$-$&$-$&$-$\\
Wilkinghoff \cite{wilkinghoff2021sub} (single model)&1&$90.7\%$&$92.8\%$&$91.8\%$&$-$&$-$&$-$&$-$&$-$&$-$\\
Wilkinghoff \cite{wilkinghoff2021sub} (ensemble)&1&$-$&$94.1\%$&$-$&$-$&$-$&$-$&$-$&$-$&$-$\\
Liu et al. \cite{liu2022anomalous}&1&$89.4\%$&$-$&$-$&$-$&$-$&$-$&$-$&$-$&$-$\\
Harada et al. \cite{harada2023first}&1&$-$&$-$&$-$&$56.9\%$&$61.1\%$&$58.9\%$&$55.4\%$&$56.5\%$&$55.9\%$\\
Wilkinghoff \cite{wilkinghoff2023design}&5&$-$&$-$&$-$&$62.8\%$&$63.0\%$&$62.9\%$&$-$&$-$&$-$\\
Hou et al. \cite{hou2023decoupling}&1&$88.8\%$&$92.0\%$&$90.4\%$&$-$&$-$&$-$&$-$&$-$&$-$\\
Wilkinghoff \cite{wilkinghoff2024self} (single model)&5&$-$&$-$&$-$&$64.2\%$&$66.6\%$&$65.4\%$&$-$&$-$&$-$\\
Wilkinghoff \cite{wilkinghoff2024self} (ensemble)&5&$-$&$-$&$-$&$-$&$70.9\%$&$-$&$-$&$-$&$-$\\
Han et al. \cite{han2024exploring}&1&$-$&$-$&$-$&$64.3\%$&$-$&$-$&$-$&$-$&$-$\\
Zhang et al. \cite{zhang2024dual}&1&$-$&$-$&$-$&$-$&$71.3\%$&$-$&$-$&$-$&$-$\\
Jiang et al. \cite{jiang2024anopatch}&1&$90.9\%$&$\pmb{94.3\%}$&$92.6\%$&$64.2\%$&$\pmb{74.2\%}$&$68.8\%$&$-$&$-$&$-$\\
Wilkinghoff \cite{wilkinghoff2024adaproj}&10&$-$&$-$&$-$&$62.9\%$&$64.5\%$&$63.7\%$&$-$&$-$&$-$\\
Saengthong et al. \cite{saengthong2024deep}&1&$74.7\%$&$-$&$-$&$-$&$73.8\%$&$-$&$-$&$-$&$-$\\
Yin et al. \cite{yin2024self}&1&$-$&$-$&$-$&$68.1\%$&$-$&$-$&$-$&$-$&$-$\\
Fujimura et al. \cite{fujimura2025improvements}&5&$-$&$-$&$-$&$67.2\%$&$68.8\%$&$67.6\%$&$-$&$-$&$62.0\%$\\
Yin et al. \cite{yin2025diffusion}&1&$-$&$-$&$-$&$-$&$-$&$-$&$-$&$\pmb{67.1\%}$&$-$\\
Jiang et al. \cite{jiang2024anopatch} presented in \cite{jiang2025adaptive}&5&$-$&$-$&$-$&$-$&$-$&$-$&$62.5\%$&$65.6\%$&$64.0\%$\\
Jiang et al. \cite{jiang2025adaptive}&5&$-$&$-$&$-$&$-$&$-$&$-$&$64.1\%$&$66.0\%$&$\pmb{65.0\%}$\\
Fujimura et al. \cite{fujimura2025asdkit}&4&$90.4\%$&$93.5\%$&$91.9\%$&$64.0\%$&$72.0\%$&$67.8\%$&$59.9\%$&$61.5\%$&$60.7\%$\\
\bottomrule
\end{NiceTabular}
\end{adjustbox}
\label{tab:sota}
\end{table*}